\documentclass[aps,amsmath,amssymb,onecolumn,notitlepage,longbibliography,floatfix,preprint]{revtex4-1}
\usepackage[dvipdfmx]{graphicx}
\usepackage{bm}

\usepackage{color,comment,ulem,here} 
\newcommand{\average}[1]{\ensuremath{\langle#1\rangle} } 
\def\R{\ensuremath{\mathrm{Re}}}
\def\Rg{\ensuremath{\R_{\rm g}}}
\def\Rt{\ensuremath{\R_{\rm t}}}
\def\Rm{\ensuremath{\R_{\rm b}}}
\def\Rc{\ensuremath{\R_{\rm c}}}
\def\Um{\ensuremath{U_{\rm m}}}
\def\Wm{\ensuremath{W_{\rm m}}}
\def\Ft{\ensuremath{F_{\rm t}}}
\def\Cf{\ensuremath{{\rm C}_{\rm f}}}

\newcommand{\rev}[1]{#1}
\newcommand{\revv}[1]{#1}
\newcommand{\revvv}[1]{#1}

\begin{document}

\title{Bifurcations to turbulence in transitional channel flow}

\author{Masaki Shimizu}
\email{shimizu@me.es.osaka-u.ac.jp}
\affiliation{%
Graduate School of Engineering Science, 
Osaka University, Toyonaka, 560-0043 Japan
}%

\author{Paul Manneville}
\email{paul.manneville@ladhyx.polytechnique.fr}
\affiliation{
Hydrodynamics Laboratory, CNRS-UMR7646, 
\'Ecole Polytechnique, Palaiseau, 91128 France
}%

\date{\today}

\begin{abstract}
In wall-bounded parallel flows\rev{,} sustained turbulence \rev{can occur} 
even while laminar flow is still stable.
Channel flow is one \rev{of} such \rev{flows} and displays spatio-temporal fluctuating patterns of localized turbulence along its way from/to featureless turbulence.
By direct numerical simulation, we study the observed inconsistency between turbulence decay according to a two-dimensional directed-percolation ($2D$-DP) scenario and the presence of sustained \rev{oblique} localized turbulent bands (LTBs) below \rev{the DP critical point}.
Above Reynolds number $\Rg \sim 700$ sustained LTBs are observed;
most LTBs \rev{have the same orientation} 
so that the spanwise symmetry of the LTB pattern is broken below $\R_2 \sim 1000$.
The frequency of transversal splitting, by which \rev{an} LTB generates another one with opposite \rev{obliqueness}, so that turbulence spreading becomes intrinsically two dimensional, increases in the range $\revv{\Rg} < \R < \R_2$.
It reaches a critical rate at $\R_2$ beyond which symmetry is restored.
$2D$-DP behavior is retrieved only above $\R_2$.
A mean-field model is proposed which qualitatively accounts for the above symmetry-restoring bifurcation by considering interactions between space-averaged densities of LTBs propagating in either \rev{direction}.
\end{abstract}

\maketitle

\section{Introduction\label{S1}}
The ways shear flows become turbulent markedly depend on the shape of their laminar velocity profiles and the occurrence of instability mechanisms of Kelvin--Helmholtz type affecting them.
Inflectional profiles like in free shear layer, wakes, or jets, become unstable at rather low values of the Reynolds number against such modes of inertial origin and then turn turbulent through a continuous cascade of successive instabilities \cite{HR98}, each step being in principle captured by weakly nonlinear perturbation theory, resulting in a {\it globally supercritical\/} transition to turbulence.
\rev{By} contrast, unidirectional flows controlled by the shear at solid walls do not display inflection points and are not prone to such instability modes.
They may become unstable but only against more \revv{intricate} and counter-intuitive mechanisms involving viscous dissipation~\cite{HR98,SH01}.
The so-produced Tollmien--Schlichting waves develop only at high Reynolds number, beyond a certain linear threshold $\Rc$ \rev{(later denoted $\R_{\rm TS}$ for convenience)}.
Pipe flow in tubes of circular and square sections or plane Couette flow between flat counter-translating plates remain linearly stable for all $\R$, $\Rc\to\infty$, whereas for channel flow (plane Poiseuille flow) $\Rc<\infty$~\cite{Or71}.
For such flows the transition is {\it globally subcritical\/}.
Physically infinitesimal perturbations are however not mathematically infinitesimal but only finite-but-small, and the possibility of a direct transition to turbulence, called {\it bypass\/}, is in practice observed at moderate values of $\R$ due to the nonlinearity of the Navier--Stokes equations:
At increasing $\R$ the flow can in principle be maintained steady and laminar up to $\Rc\le \infty$, but the introduction of finite perturbations with particular shapes and amplitudes may force it to become unsteady.
There is therefore a full range of Reynolds numbers, called {\it transitional\/}, where the trivial laminar solution coexists with other nontrivial solutions.
These solutions are usually chaotic and cannot be straightforwardly reached by standard perturbation analysis.
A direct transition to turbulence is observed with strong hysteresis  upon continuous up-and-down variation of $\R$, the flow jumping from the laminar solution branch to the nontrivial turbulent branch or back.
Once in a state belonging to the nontrivial branch, the flow can be maintained  turbulent down to a global stability threshold $\Rg$ below which the only possible stable regime is laminar.

Sub-criticality is often perceived as a problem to be studied within the theory of dynamical systems.
Solutions to the Navier--Stokes equations are then searched under minimal flow unit (MFU) conditions~\cite{JM91} with periodic boundary conditions at short distances, focusing on coherent structures analyzed in the corresponding {\it phase space\/} perspective~\cite{Wa01,K12}.
This approach was progressively generalized to deal with extended systems in one or two in-plane directions, and mostly concentrated on the existence and properties of isolated localized solutions at the edge of chaos, see e.g.~\cite{ZE14} for channel flow.
On the other hand, when appreciated in {\it physical space\/}, provided that the system's geometry is wide enough, states along the nontrivial branch in general display a coexistence of domains alternately laminar and turbulent separated by fluctuating interfaces.

In a number of cases, beyond $\Rg$ this laminar-turbulent coexistence happens in the form of a regular pattern.
The iconic case is that of the turbulent spiral obtained in moderate-gap cylindrical Couette flow by Coles~\cite{Co65}, and re-obtained later by Andereck {\it et al.}~\cite{An86} who introduced the term {\it featureless\/} to qualify the regime obtained at higher shearing rates when turbulence has apparently returned to statistical uniformity.
In small-gap experiments, several helical branches were obtained~\cite{P03}.
Close to the featureless regime, laminar-turbulent patterning involved superpositions of such helices with opposite pitch while, plane Couette flow being understood as the zero-curvature limit of cylindrical Couette flow, helices were straightened into oblique bands found in a whole range of  $\R$ down to $\Rg$ \cite{P03}.
As $\R$ increased, the amplitude of the turbulence modulation associated with these bands/helices was seen to decrease, and a threshold usually denoted as $\Rt$ could be determined above which turbulence was featureless.
The transition at $\Rt$ seemed continuous with a major role played by the intense noise arising from the chaotic dynamics in the turbulent regime~\cite{P03,TB11}.
Regular laminar-turbulent patterning with similar characteristics has been observed in a few other cases including stratified Ekman boundary layer~\cite{De14}, Couette--Poiseuille flow~\cite{Kl17}, or stratified plane Couette flow~\cite{De15}.
Experimental evidence for laminar-turbulent oblique patterning in channel flow was provided by Tsukahara's group~\cite{T09} as a follow-up of their simulations in wide domains~\cite{T05}.
Another numerical approach using the oblique-elongated-but-narrow domain assumption~\cite{BT05}, was developed in~\cite{Tu14} to yield similar patterns at a much more limited numerical cost.

The location of the lower threshold $\Rg$ is another problem that can be studied either by increasing $\R$ from a germ or \revv{by} 
decreasing $\R$ from a pattern.
In the first case the study mostly relies on the capability to produce localized states and study their persistence.
The first solid results were obtained in pipe flow (consult~\cite{E08} for a general perspective and references back to Reynolds' seminal work) for which the relevant localized states are {\it puffs\/} and the threshold $\Rg$ identified the value of $\R$ when the probability of turbulence extinction due to their decay is overcome by the probability of their proliferation due to repeated splittings~\cite{A11}.   
In planar flows, such as in planar boundary layers, plane Couette flow~\cite{TA92}, Couette--Poiseuille flow~\cite{Kl17}, or channel flow~\cite{CWP82}, equivalent localized solutions appear as {\it turbulent spots} (see~\cite{LAW13} for a recent study).
\rev{The specific effect of spanwise extension on localization has been studied in small aspect-ratio rectangular duct flow~\cite{Tetal15} and in annular Poiseuille flow~\cite{Ietal17}.}
In all cases the location of $\Rg$ is derived from statistical studies, with the difficulty that a wide range of initial conditions needs to be considered by varying their shapes, structures and strengths.
A value $\Rg\sim 1000$ was mentioned in~\cite{CWP82} for channel flow, and was also reported in other systems provided that the Reynolds number is based on an appropriate physically-based estimate of the shear~\cite[Appendix]{BT07}.
The second approach, i.e., by decreasing $\R$, is also statistical but rather appeals to an early conjecture by Pomeau~\cite{P86} who put forward the analogy between turbulence onset/decay in extended systems and directed percolation (DP).
This conjecture was based on the recognition that DP is a spatiotemporally intermittent process involving {\it active\/} (here turbulent) and {\it absorbing\/}  (here laminar) states defined locally in physical space, and the probability of contamination of the \revv{latter}  
by the \revv{former} is the key factor.
DP is the representative of a class of non-equilibrium phase transition with specific universal properties depending on the effective dimension $D$ of the physical space in which the process develop \cite{H08}.
Conceptually, the corresponding critical point $\R_{\rm DP}$ should have to be identified with $\Rg$ and the long turbulent transients observed for $\R<\Rg$ as the decay of a percolation cluster below threshold.

Flows evolve physically in three-dimensional domains but confinement in the wall-normal direction reduces the dimension to an effective value $D=1$ in tubes and flow configurations similarly constrained by lateral boundary conditions, or $D=2$ when the lateral boundaries are far enough when compared to the typical wall-normal distance.
This is due to the fact that the transitional range takes place at moderate $\R$ so that viscosity is able to impose a strong coherence except along the unbounded directions, one in tubes or two along plates.
To our knowledge, $1D$-DP universality has not been checked directly in tubes up to now, but a numerical model suggests it might apply~\cite{Ba11}.
On the other hand, universality has been shown to hold quantitatively in a numerical model of plane shear flow defined in a narrow oblique domain~\cite{S13,L16}  (a geometry introduced in \cite{BT05}) and in a cylindrical Couette configuration with a small axial aspect ratio ensuring quasi-$1D$ confinement~\cite{L16}.
Similar findings have also been obtained in a model of two-dimensional shear flow without walls called Kolmogorov flow in an elongated domain rendering the dynamics effectively $1D$~\cite{hiruta2018subcritical}.
As to $2D$-DP universality, the corresponding critical behavior has been obtained with great precision in a numerical experiment on a low order Galerkin approximation to Waleffe flow, which is a three-dimensional shear flow mimicking plane Couette flow but with stress-free boundary conditions~\cite{C17}.
Preliminary results on plane Couette flow proper, but in an under-resolved numerical context~\cite{Sh17}, suggest a similar conclusion.
At last, agreement with $2D$-DP universality has been obtained by Sano \& Tamai~\cite{ST16} in a laboratory experiment on channel flow through a duct with a rectangular section and large spanwise aspect ratio.
This finding has however to be re-interpreted in view of the discovery of sustained nontrivial solutions below $\R_{\rm DP}$ that could therefore not be the expected global stability threshold.

These solutions were first obtained in the form of a localized turbulent band (LTB) by triggering~\cite{Tao13} and later upon looking for the global stability threshold by slowly decreasing $\R$ and following the mutations in the laminar-turbulent patterning~\cite{X15,T15}, \rev{by} contrast with experiments in~\cite{ST16} where the decay from homogeneous turbulence was studied in a channel of limited length.
They were later studied in more detail in~\cite{K17} where sustainment mechanisms were scrutinized and in \cite{T18} where their decay was considered in relation with the size of the numerical domain.
Localized solutions with an analogous structure were also obtained in laboratory experiments~\cite{H16,Pa17}.

Below we present our simulation results for channel flow in wide \revv{domains}.
An overview of the observed flow regimes is given in \S\ref{S2}, from LTBs around $\Rg$ to the oblique pattern regime described in \cite{T05,T09} and to featureless turbulence above $\Rt$.
We next show in two steps how the contradiction between the observation of a $2D$-DP scenario as reported in \cite{ST16} can be reconciled with that of LTBs at lower $\R$.
In \S\ref{S3}, we enter the details of how LTBs grow and split as $\R$ increases, and how spanwise symmetry is restored from the low-$\R$ regime where LTBs move essentially in one direction~\cite{Tao13,X15,T15,K17,T18,H16,Pa17}.
A simple phenomenological model is developed to show that this bifurcation is controlled by the increasing rate of {\it transversal\/} splitting, thereby  supporting the existence of a well-defined threshold.
Above that threshold the invasion of turbulence is genuinely $2D$ in the plane of the flow, in the form of a spatiotemporally intermittent network, where a DP-like behavior becomes relevant.
The quantitative analysis of the turbulent fraction in \S\ref{S4} will give evidence of a growth compatible with $2D$-DP universality.
Our findings are recapitulated in~\S\ref{S5} where we discuss in particular how processes specific to particular flow configurations may come and breach the appealing concept of universality expected on general grounds.  
Details on our simulations and the filtering-thresholding methodology used to measure the turbulent fraction are presented in Appendices~\ref{AppA} and~\ref{AppB}, respectively.

\section{Overview of the transitional range\label{S2}}

Channel flow here is driven by a constant body force~$f$. 
The Reynolds number is defined as $\R\!=\!fh^3/2\nu^2$, where $h$ is the half-distance between two parallel walls and $\nu$ the kinematic viscosity.
All quantities below are written in units of the center-plane velocity of the corresponding laminar flow $U\!=\!fh^2/2\nu$ and $h$.
For comparison with previous works, we also define $\Rm = \frac32 \average{\Um} \R \,$, where $\average{\Um}$ is the time-average of the dimensionless bulk velocity $\Um$ (`b' stands for `bulk'  \cite{Tu14}).
\revv{Most of the time, the computational domain size is $500 \times 250$ (streamwise $\times$ spanwise) but we also consider $250 \times 125$ and $1000 \times 500$ to check for size effects. 
Simulations are performed for durations sufficient to obtain statistically significant results, typically up to $\revvv{1.5 \times 10^5}$ time units.}
See Appendix \ref{AppA} for a detailed description of the flow system and the numerical procedures.

\begin{figure}[!b]
\includegraphics[width=\textwidth]{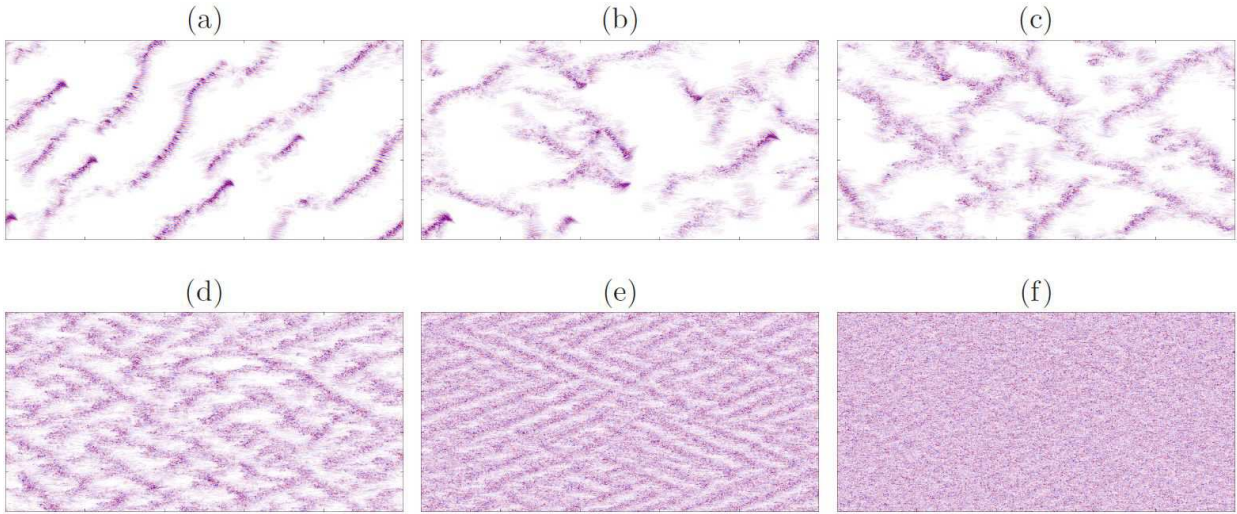}
\caption{
(a) $\R=850$ ($\Rm\approx789$): one-sided LTB regime. 
(b) $\R=1050$ ($\Rm\approx947$): two-sided LTB regime. 
(c) $\R=1200$ ($\Rm\approx1012$): 
strongly intermittent loose continuous network of LTBs. 
(d) $\R=1800$ ($\Rm\approx1237$): weakly intermittent loose banded pattern. 
(e) $\R=3000$ ($\Rm\approx1604$): tight banded pattern. 
(f) $\R=4000$ ($\Rm\approx1843$): nearly featureless state.
Flow direction is from the left to right. 
The wall-normal velocity field on the center plane $(y=0)$ is displayed. \rev{Domain size is $500\times250$.}
See Supplemental Videos~\cite{SM}. 
\label{snap}}
\end{figure}

\begin{figure*}
\includegraphics[width=0.7\textwidth]{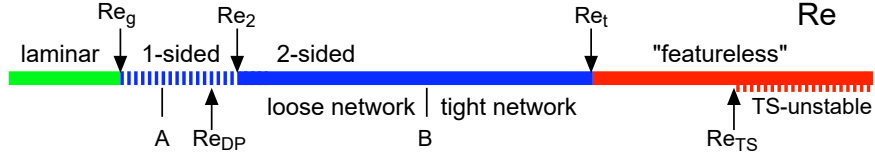}
\caption{\label{sketch} \rev{Bifurcation diagram for plane channel flow as obtained in our simulations.
(Corresponding values of $\Rm$ between parentheses.) Laminar flow is always recovered in the long-time limit for $\R<\Rg \approx 700$ ($\R\equiv\Rm$ up to $\Rg$). Event A corresponds to the onset of transversal splitting at $\R \sim 800$ ($\Rm \sim 795$). The extrapolated $2D$-DP threshold is found at $\R_{\rm DP} 
\simeq 984$ ($\Rm \simeq 905$).
The transition from one-sided to two-sided propagation of LTBs takes place at $\R_2 \simeq 1011$ ($\Rm \simeq 924$).
Event B marks the opening of laminar gaps in the turbulent branches, sufficiently long-lived to allow the development of DAHs. It takes place at  $\R\approx1200$ ($\Rm \approx 1012$). Finally, ``featureless'' turbulence is present for $R > \Rt \approx 3900$ ($\Rm\approx 1820$). 
The Tollmien--Schlichting instability threshold is at $\R_{\rm TS}=5772$ (analysis refers to laminar base flow, hence identical $\Rm$).} }
\end{figure*}

Snapshots of flow patterns for typical Reynolds numbers are displayed in Fig.~\ref{snap}.
For $\R=850$ and $1050$, Fig.~\ref{snap}(a,b), several localized turbulent bands (LTBs)  are observed, propagating at an angle about $45^\circ$ with the streamwise direction, each driven by a downstream active head (DAH)~\cite{Tao13,X15,T15,T18,TT16,K17} located at its downstream extremity.
DAHs \rev{entraining LTBs} drift at a speed about 0.8 in the streamwise direction and about 0.1 in the spanwise direction.
In agreement with previous studies  \cite{K17,T18}, these LTBs decay below $\R_{\rm g}\approx700$.  
At $\R=850$~(a), all \rev{LTBs} go in the same direction, therefore breaking the symmetry with respect to the spanwise direction.
\rev{By} contrast, \rev{LTBs} go in both directions at $\R=1050$~(b).
These states are respectively called one-sided and two-sided. 
As $\R$ increases, LTBs joint to form a loose continuous network of oblique bands and for $\R=1200$~(c) DAHs practically cease to be seen.
The pattern is strongly intermittent with turbulence intensity far from being uniform along the bands.
At larger values of $\R$, the network narrows, $\R=1800$~(d), and wide laminar voids disappear while regular patterns form, which can be understood as crisscrossed more acute ($\sim 25^\circ$) oblique turbulence modulations, $\R=3000$~(e), similar to those obtained in circular Couette flow~\cite{P03}.
The amplitude of this modulation then decreases and the {\it featureless\/} regime eventually prevails for $\R\gtrsim 4000$~(f).
Properties of the flow deeper inside the uniformly turbulent, developed regime achieved at even larger values of $\R$ are briefly reviewed in \cite{Pp00}.
\rev{Figure~\ref{sketch} is a sketch of the bifurcation diagram of channel flow indicating the different regimes illustrated above and anticipating the output of the quantitative study of phenomena developing for $800\le \R \le 1200$ to be presented below.}

\begin{figure}[t]
\includegraphics[width=\textwidth]{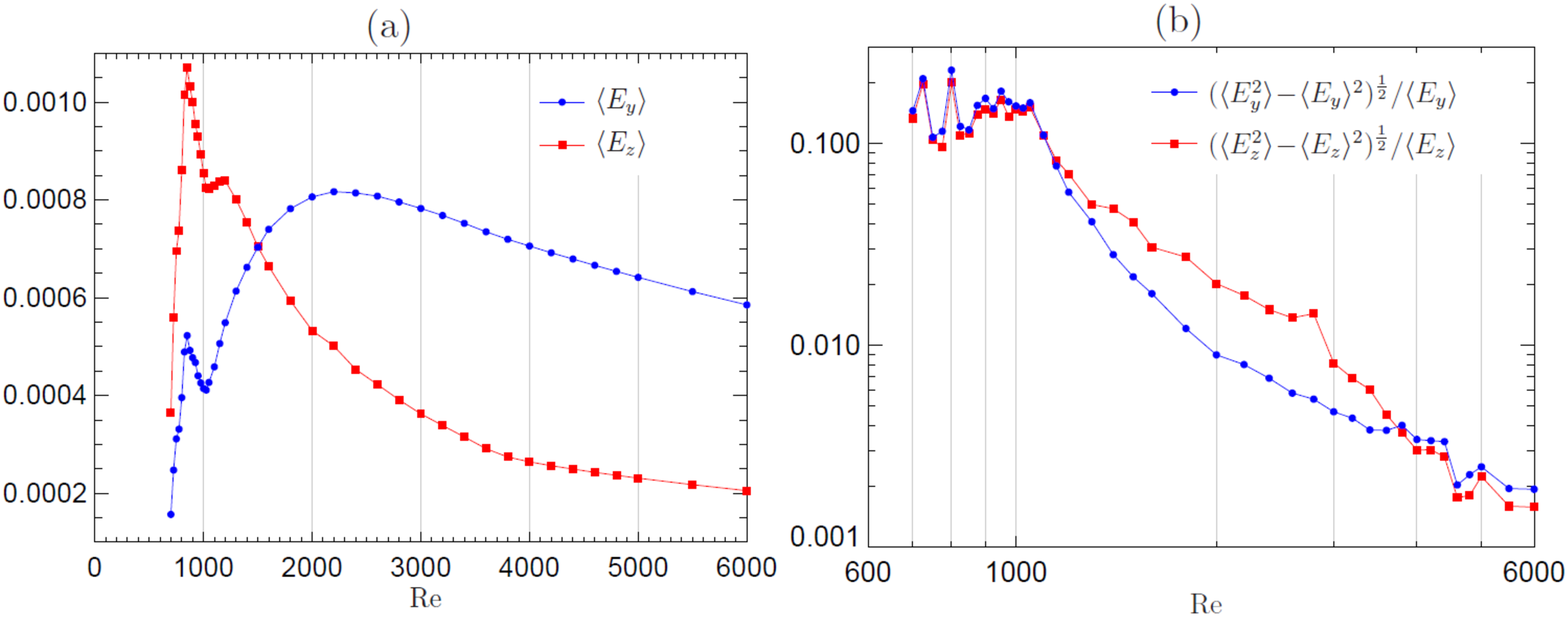}
\caption{\label{mean}
Means (a) and relative standard deviations (b) of the wall-normal energy $E_y(t)$ and spanwise energy $E_z(t)$.}
\end{figure}

\begin{figure*}[t]
\includegraphics[width=\textwidth]{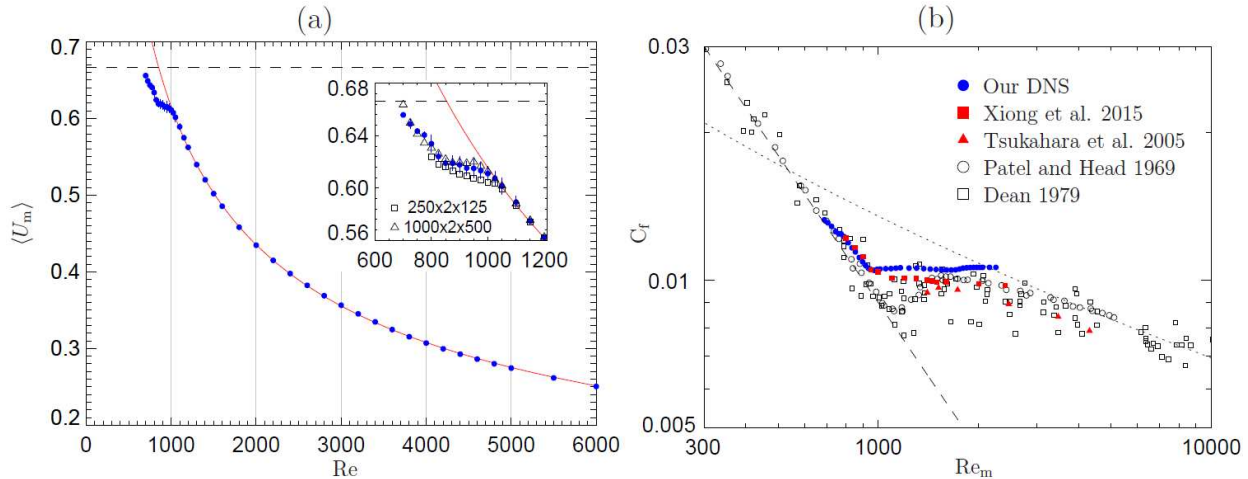}
\caption{
(a) Average of streamwise bulk velocity $\average{\Um}$ as a function of $\R$; the solid line corresponds to the fit and the dashed line to laminar flow, $\Um^{\rm lam}=2/3$. 
Open symbols in the inset are for the different domain sizes.
(b) Skin friction coefficient ${\rm C}_{\rm f}$ as a function $\Rm$. Filled and open symbols 
are from DNSs \revvv{(Tsukahara et al.~2005~\cite{T05}, Xiong et al.~2015~\cite{X15})} 
and experiments \revvv{(Patel and Head~1969~\cite{PH69}, Dean 1979~\cite{D78})}, respectively.
The dashed line is for laminar flow and the dotted line for the fully turbulent regime.
\label{mean2}}
\end{figure*}

Information from the statistics over the time-series of typical global quantities is displayed as functions of $\R$ or $\Rm$ in Fig.~\ref{mean} and \ref{mean2}.
Transverse turbulent energies, $E_y(t)={\mathcal V}^{-1} \int_{\mathcal V} u_y^2 {\rm d} {\mathcal V}$ and $E_z(t)={\mathcal V}^{-1} \int_{\mathcal V} u_z^2 {\rm d} {\mathcal V}$,  in Fig.~\ref{mean}(a) directly monitor the distance to the laminar base flow.
Irregularities noted for $\R<1200$ can be interpreted with the help of Fig.~\ref{snap}(a--c).
The rapid growth of $E_y$ and $E_z$ for $\R\le 850$ is related to the increasing number of DAHs in the one-sided LTB regime.
When $\R>850$, the increasing fraction of LTBs with different orientations leads to a strong decrease in $E_y$ and $E_z$ until $\R \approx 1000$. 
Next, as $\R$ increases, $E_y$ grows again owing to an increasing turbulent fraction while $E_z$ slightly increases up to $\R \approx 1200$ before decreasing  in the band-network regime where the global flow around LTBs is inhibited.
The variations of the standard deviations of fluctuations, once normalized by their respective means, are remarkably correlated, as seen in Fig.~\ref{mean}(b).
Their rapid growth as $\R$ approaches \revv{$\R \approx 1000$} from above
is reminiscent of the divergence of fluctuations observed for a phase transition at a threshold $\R_2$ precisely located later.
Peaks at $\R=725$ and $\R=800$ mark the onset of longitudinal and transversal splittings to be examined below. 

Further general observations about the transitional range as a whole can be extracted from our numerical results.
Using a constant body force (mean applied pressure gradient), our numerical implementation of Navier--Stokes equations produces streamwise and spanwise net flux components $\Um$ and $\Wm$ as time-fluctuating observables governed by (\ref{e4}).
In  particular, the bulk Reynolds number $\R_{\rm b}:=\frac32 \average{\Um}\R$ introduced earlier is obtained from the time average of $\Um$.
Figure~\ref{mean2}(a) displays $\average{\Um}$ as a function of $\R$.
Laminar flow corresponds to $\Um^{\rm lam}=2/3$.
As soon as some turbulence is present we get $\average{\Um}<\Um^{\rm lam}$ and, for $\R$ above the one-sided regime, the observed decrease nicely fits an inverse square root over a large part of the transitional range.
Using results for $\R\ge1050$  we obtain $\Um^{\rm fit}=w/\sqrt{\R}$ with $w\simeq19.5$, hence $\Rm \approx  \frac32 w \sqrt{\R}$.
Turning to wall units, the friction velocity $U_\tau$ and the friction Reynolds numbers $\R_\tau$ are obtained  in our formulation as $U_\tau^2= \tau_{\rm w} = 2/\R$ and $\R_\tau=\sqrt{2\R}$, relation (\ref{Retau}).
This means that $\R_\tau$ and $\Rm$ are roughly proportional as long as the flow remains textured, before entering the developed regime where $\R_\tau$ then grows as a function of $\Rm$ at a slightly smaller rate (exponent $\simeq0.88$~\cite{Pp00}).
This behavior is directly reflected in the variation of the skin friction coefficient $\Cf$  with $\Rm$ shown in Fig.~\ref{mean2}(b).
 Here we have $\Cf:=\tau_{\rm w}/\frac12\average{\Um}^2=4/(\R\,\average{\Um}^2)$ (see  \S\ref{AppA}).
 As soon as some turbulence is present, $\Cf$ deviates from its laminar expression $9/\Rm$, remains close to it in the one-sided regime, next change to a near plateau dependence as soon as it enters in the two-sided regime, as expected from the variation of $\average{\Um}^2$ as $\R^{-1}$ in the corresponding $\R$-range.
In our simulations, this plateau  extends up to the transition to fully developed turbulent channel flow marked by a dotted line in Fig.~\ref{mean2}(b), theoretically expressed as $\Rm=\frac{3}{\sqrt{2\Cf}}\, \exp\big(0.41\big(\sqrt{2/\Cf}
-2.4\big)\big)$~\cite{D78,Pp00}.
This plateau is better defined than in earlier experiments~\cite{PH69,D78} and simulations \cite{X15,T05} also shown in the picture.
It indicates that, above $\R_2$, the transition to turbulence develops at a nearly constant dissipation rate $\average{\epsilon_{\rm m}}$, as stems from relation (\ref{epsilonm}).
\rev{In the upper transitional range $\R\gtrsim4000$, i.e. $\Rm\gtrsim1850$, the trend suggested by our data points seems to overestimate $\Cf$ as predicted for the fully turbulent regime~\cite{Pp00} (dotted line).
This is presumably due to a lack of resolution close to the wall at such high values of the Reynolds number.
A detailed quantitative study of the transition to the ``featureless'' regime around $\Rt$, beyond the preliminary result in~\cite{MS19} indicated in Figure \ref{sketch}, is left to future work.}

\section{Symmetry-restoring bifurcation\label{S3}}

\begin{figure*}[t]
\begin{minipage}{0.32\hsize}
(a1)\hspace*{3ex}$\R=725$, $t$=8000  \\
\includegraphics[width=\textwidth]{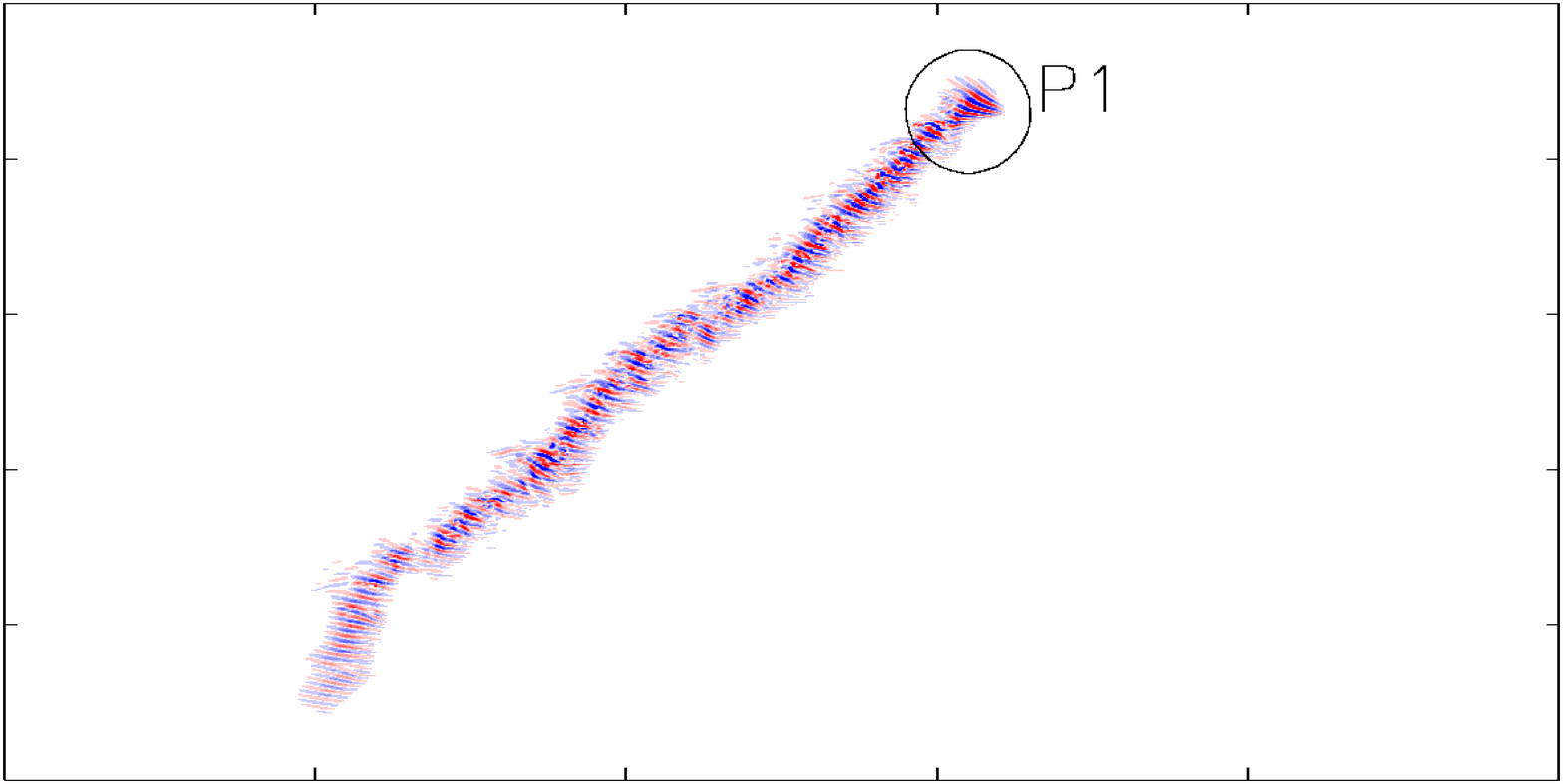}
(b1)\hspace*{3ex}$\R=725$, $t$=59000 \\
\includegraphics[width=\textwidth]{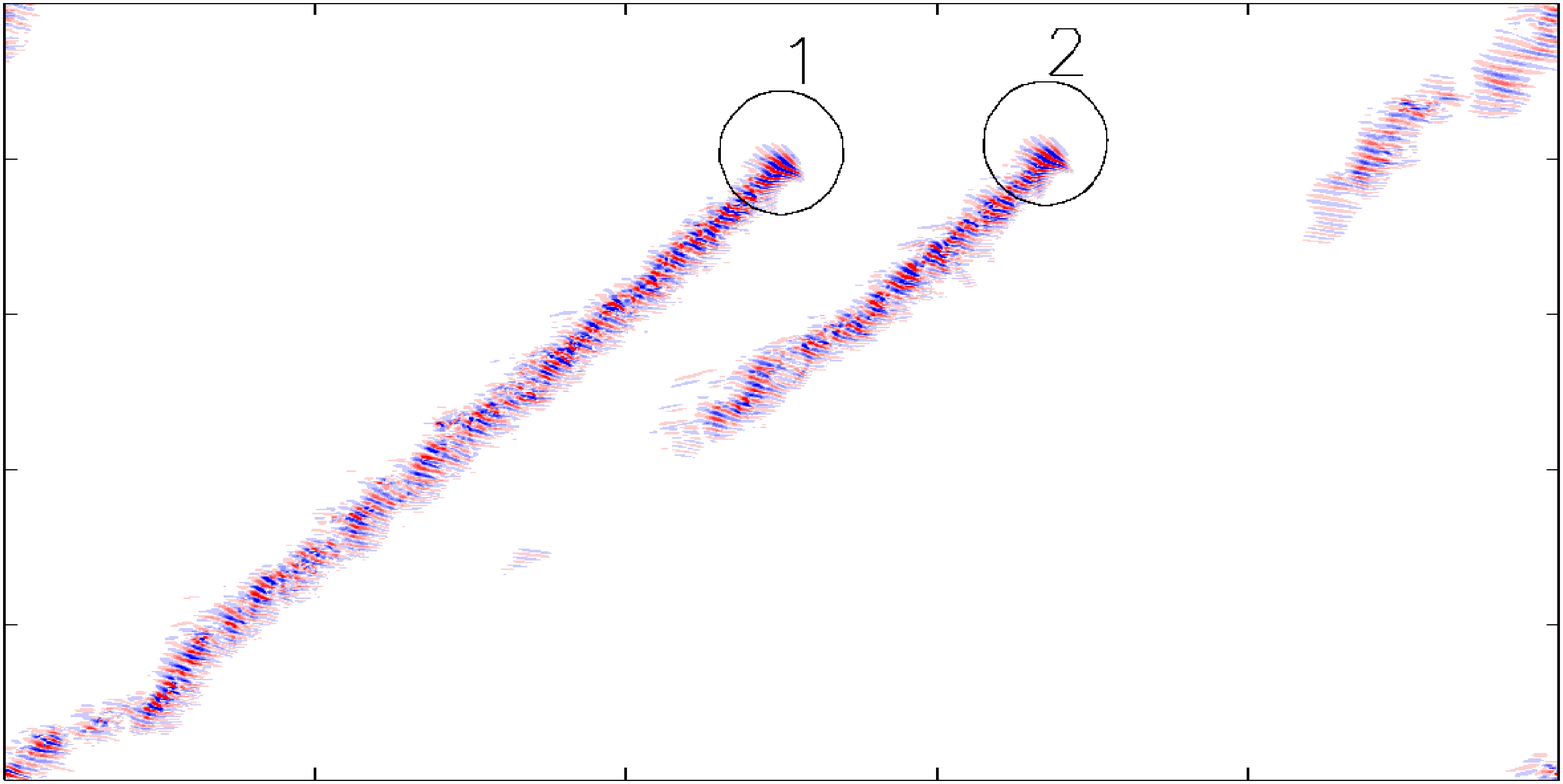}
(c1)\hspace*{3ex}$\R=900$, $t$=53200 \\
\includegraphics[width=\textwidth]{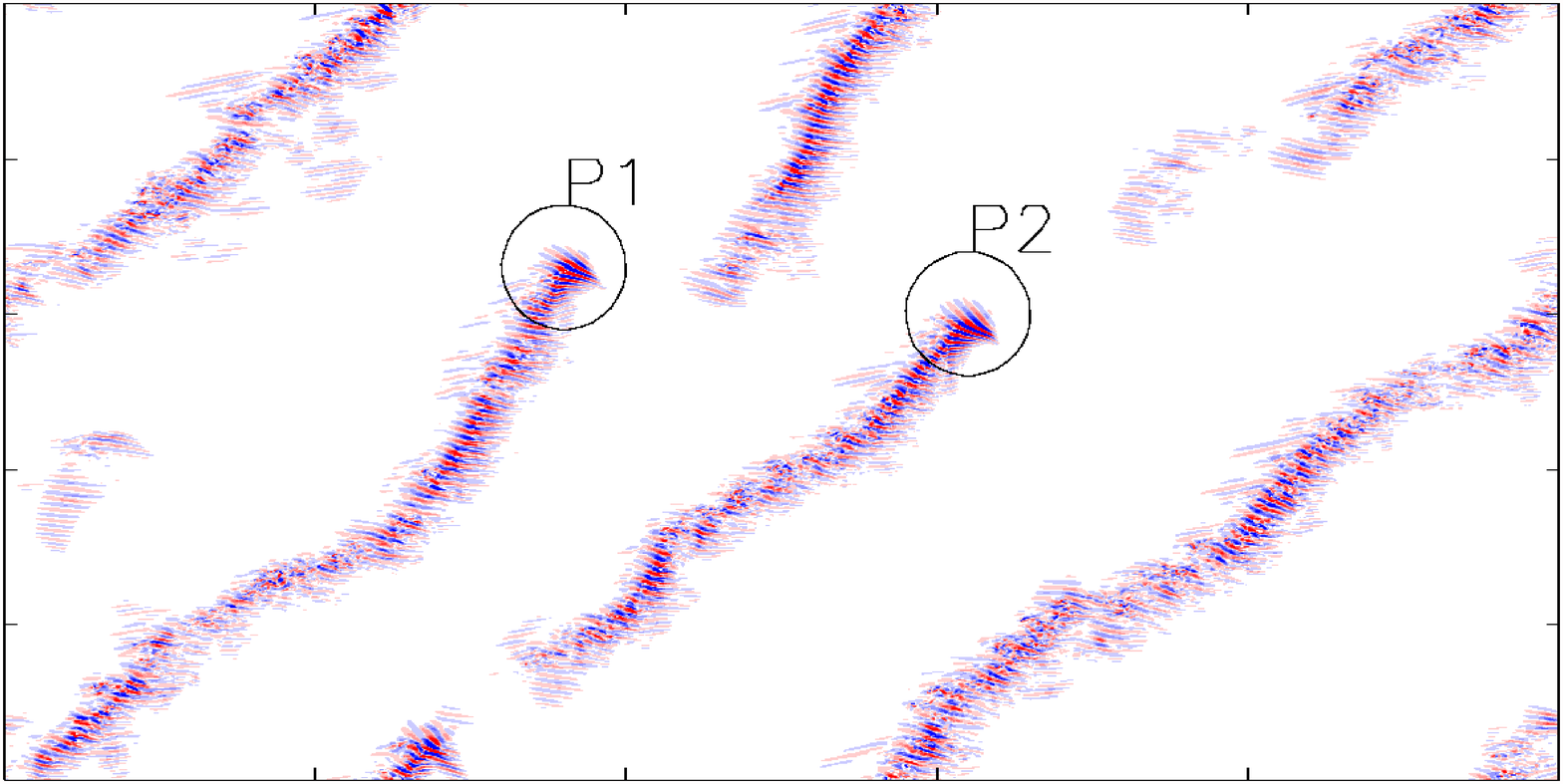}
(d1)\hspace*{3ex}$\R=900$, $t$=136900\\
\includegraphics[width=\textwidth]{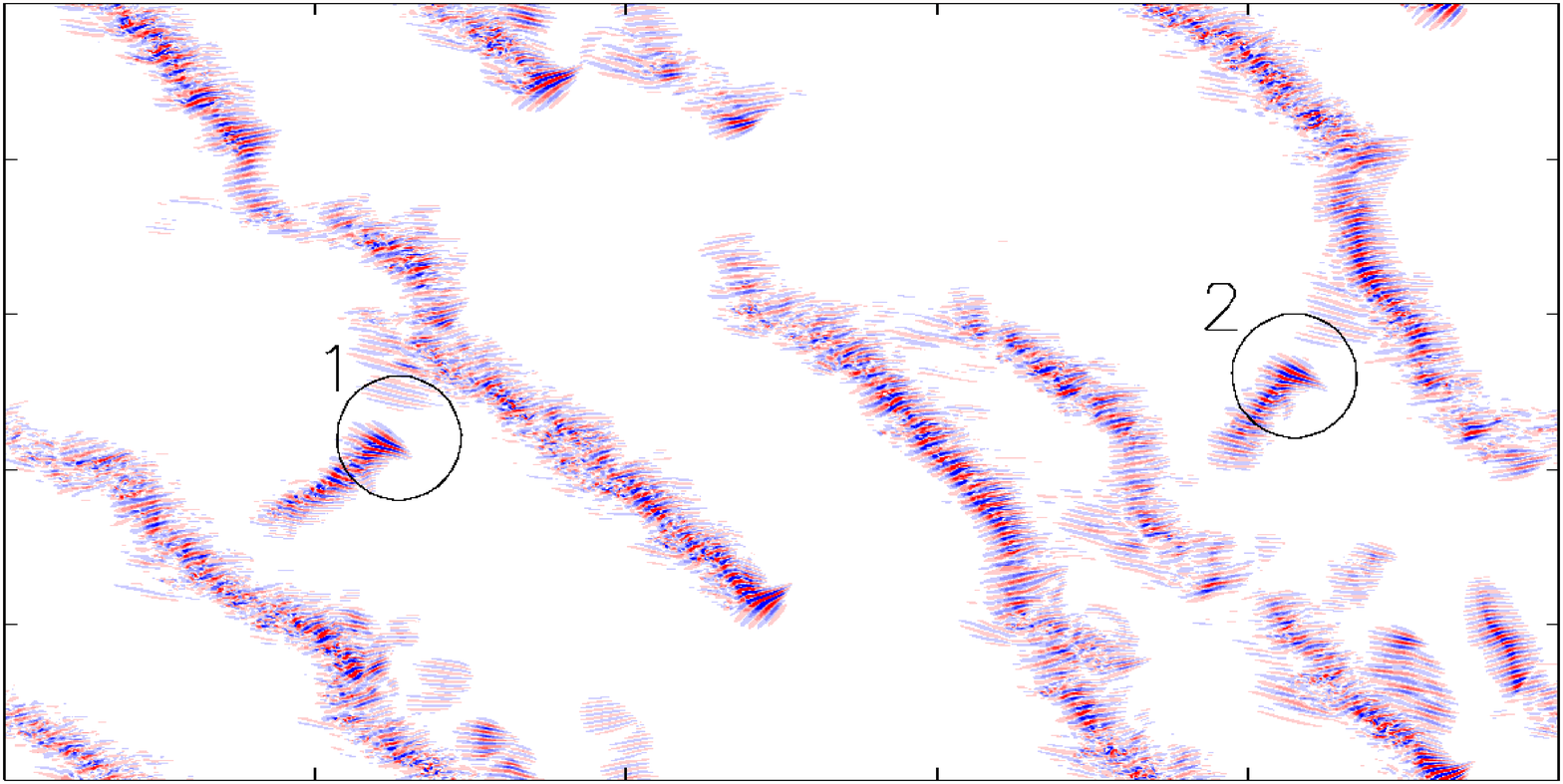}
\end{minipage}\hfill			
\begin{minipage}{0.32\hsize}
(a2)\hspace*{3ex}$\R=725$, $t$=9000  \\
\includegraphics[width=\textwidth]{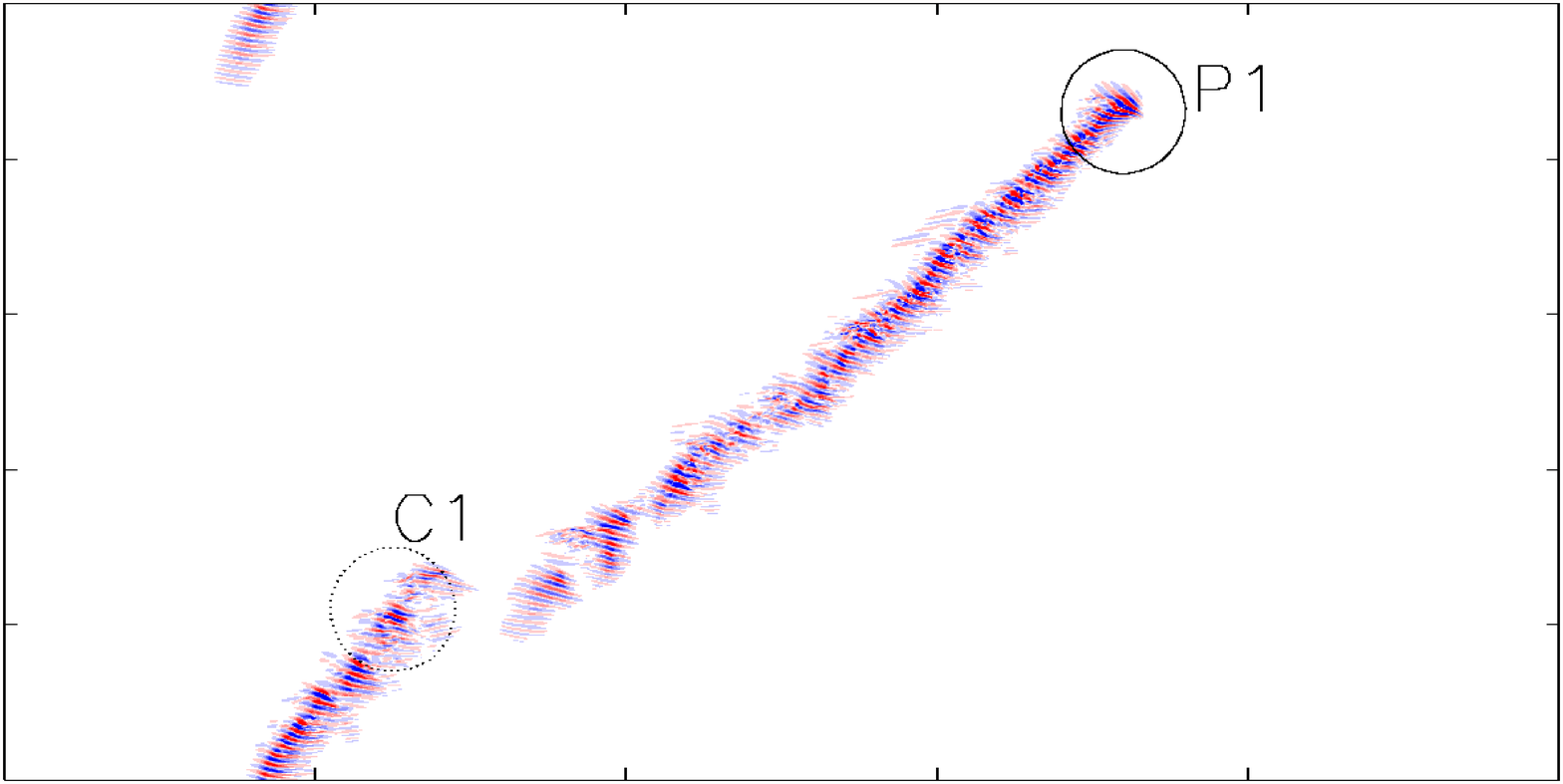}
(b2)\hspace*{3ex}$\R=725$, $t$=62500 \\
\includegraphics[width=\textwidth]{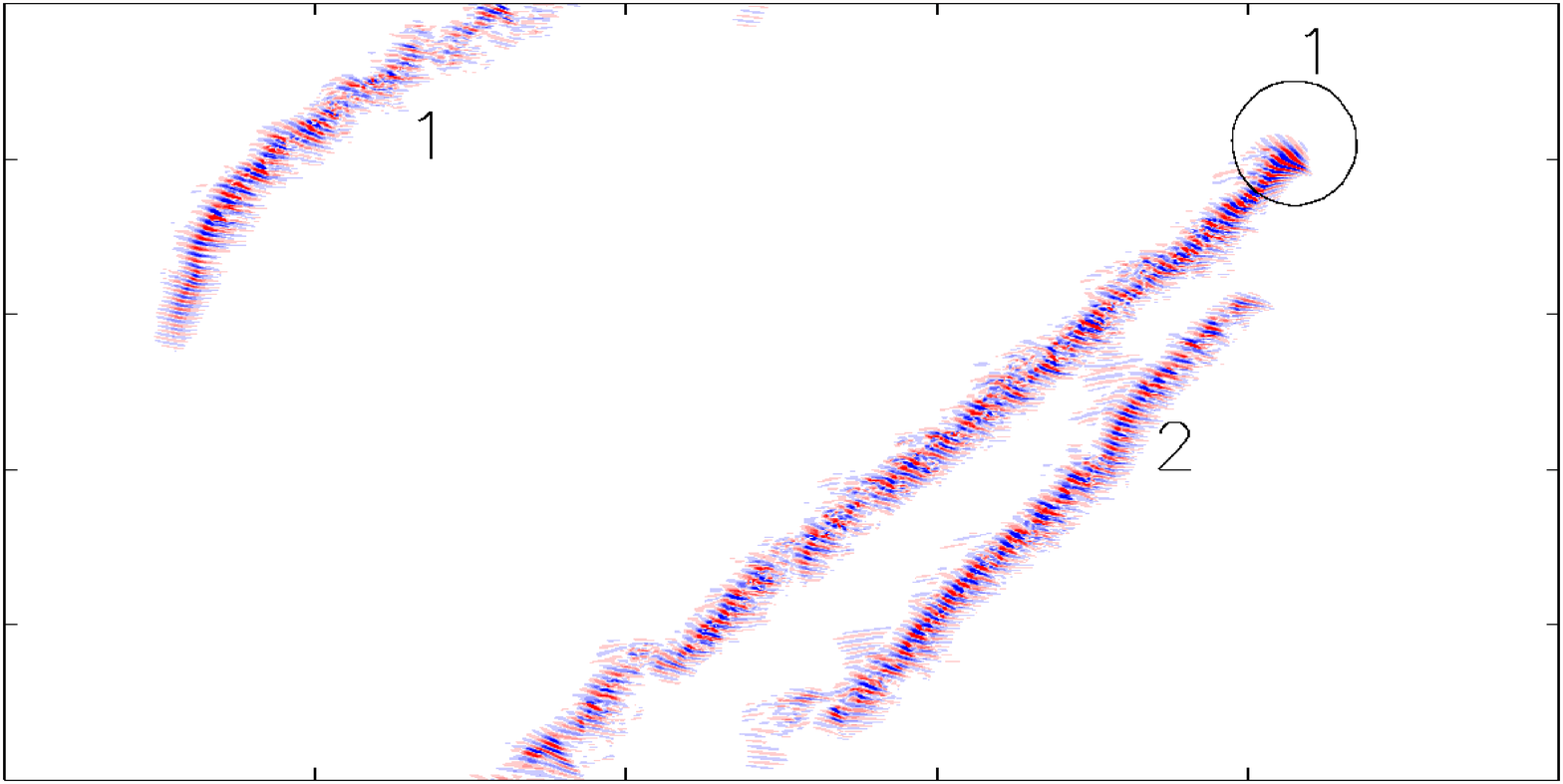}
(c2)\hspace*{3ex}$\R=900$, $t$=53500 \\
\includegraphics[width=\textwidth]{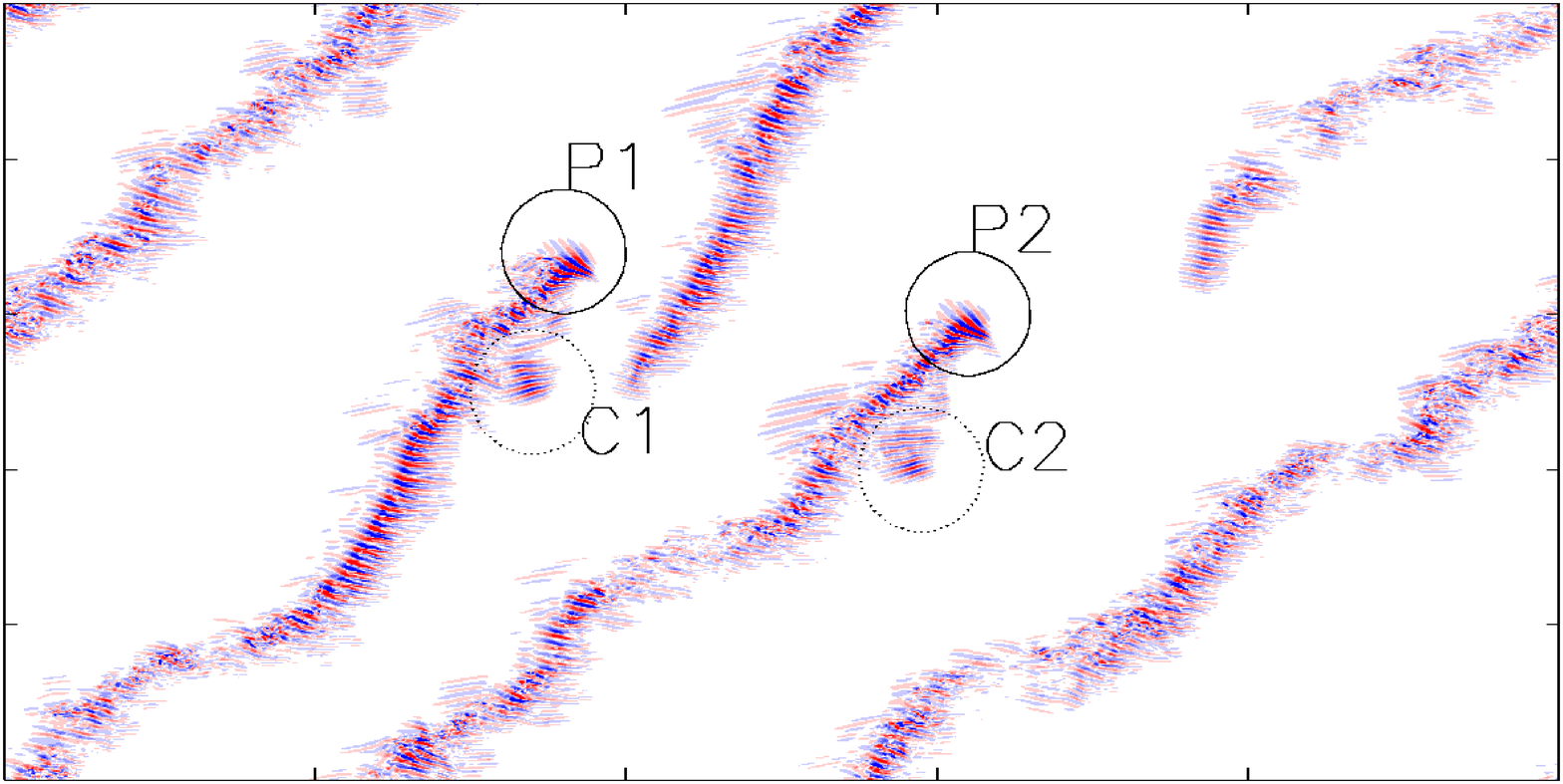}
(d2)\hspace*{3ex}$\R=900$, $t$=137100\\
\includegraphics[width=\textwidth]{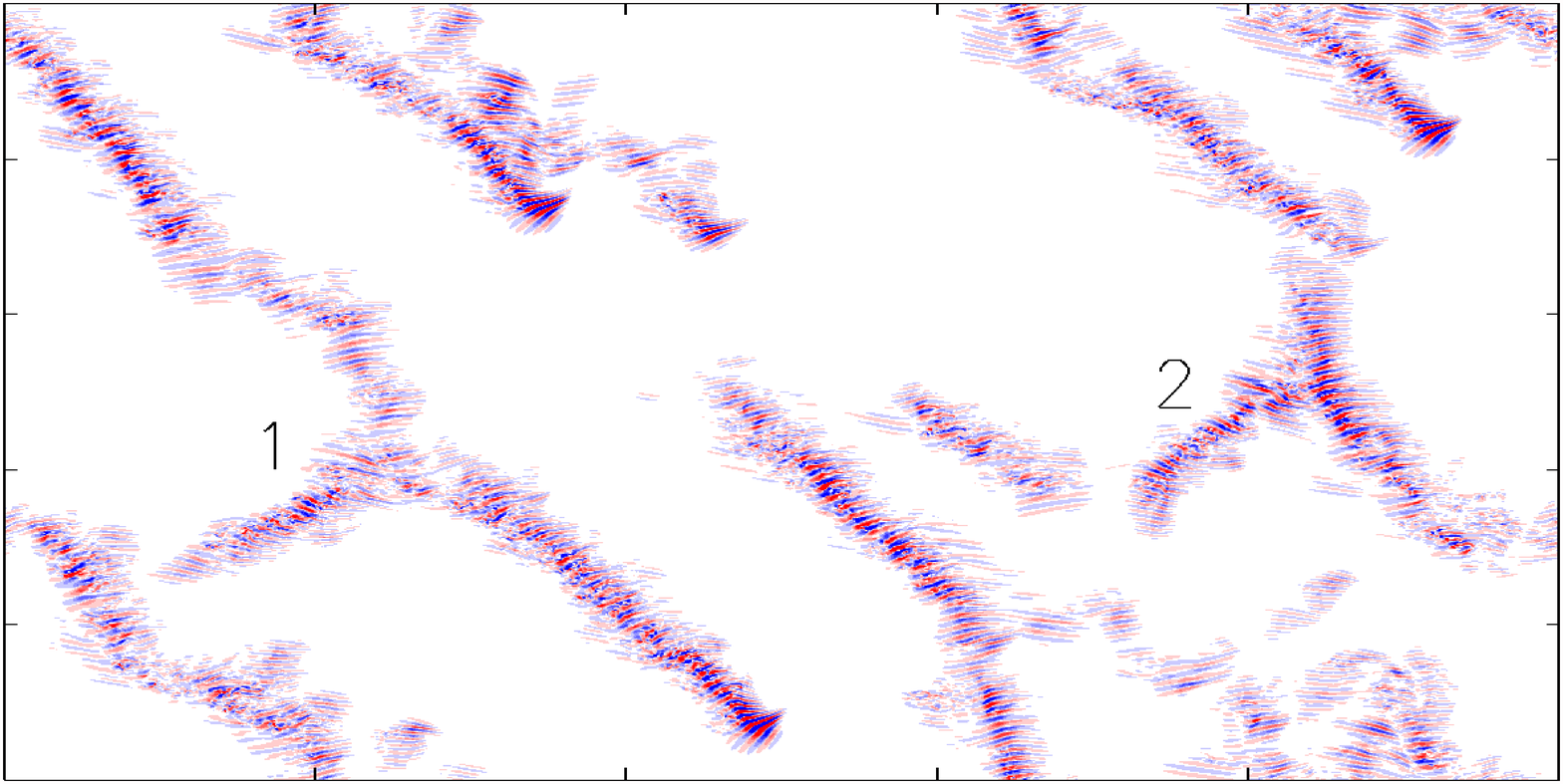}
\end{minipage}\hfill
\begin{minipage}{0.32\hsize}
(a3)\hspace*{3ex}$\R=725$, $t$=10000 \\
\includegraphics[width=\textwidth]{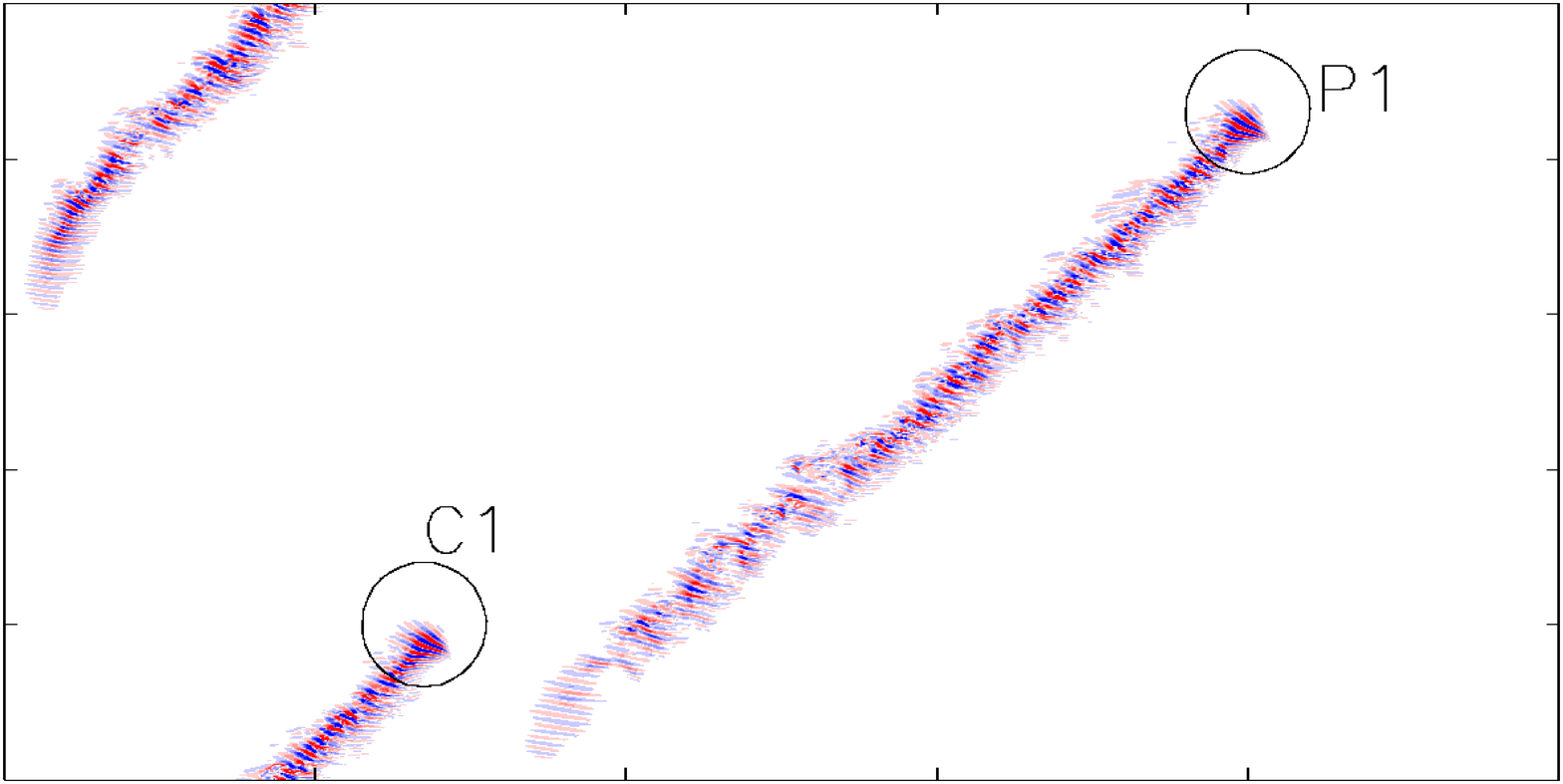}
(b3)\hspace*{3ex}$\R=725$, $t$=63500 \\
\includegraphics[width=\textwidth]{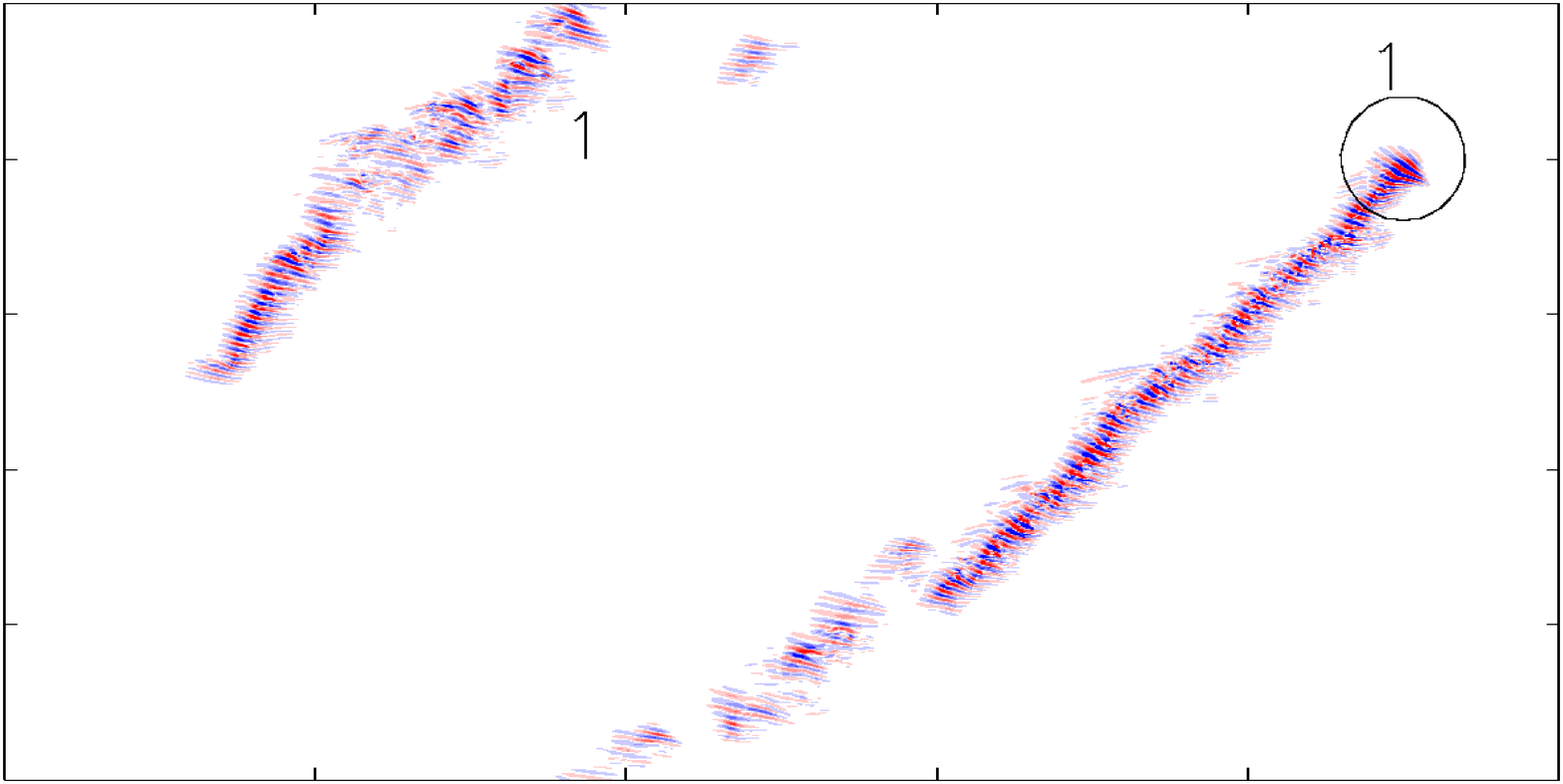}
(c3)\hspace*{3ex}$\R=900$, $t$=53800 \\
\includegraphics[width=\textwidth]{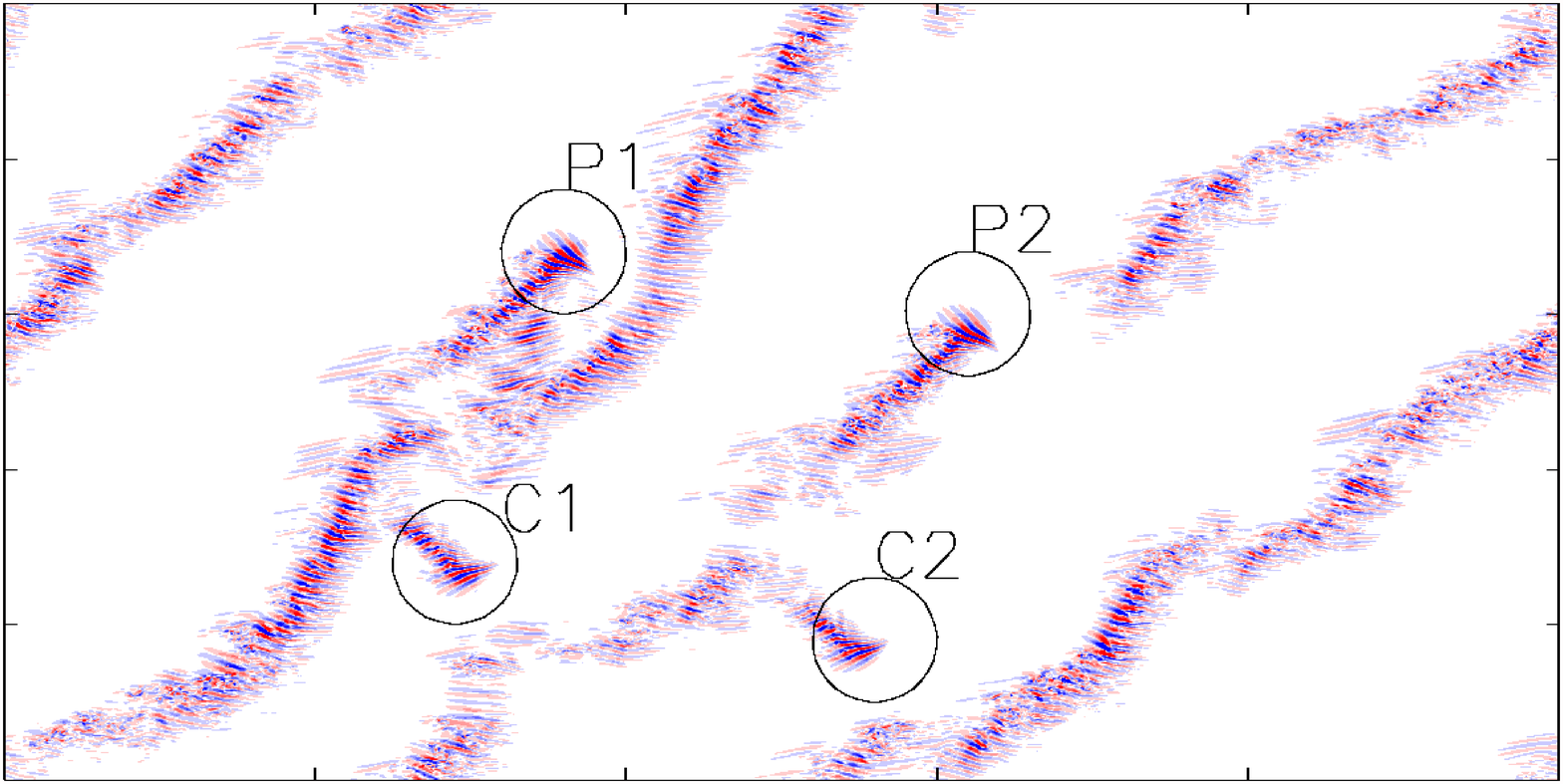}
(d3)\hspace*{3ex}$\R=900$, $t$=137400 \\
\includegraphics[width=\textwidth]{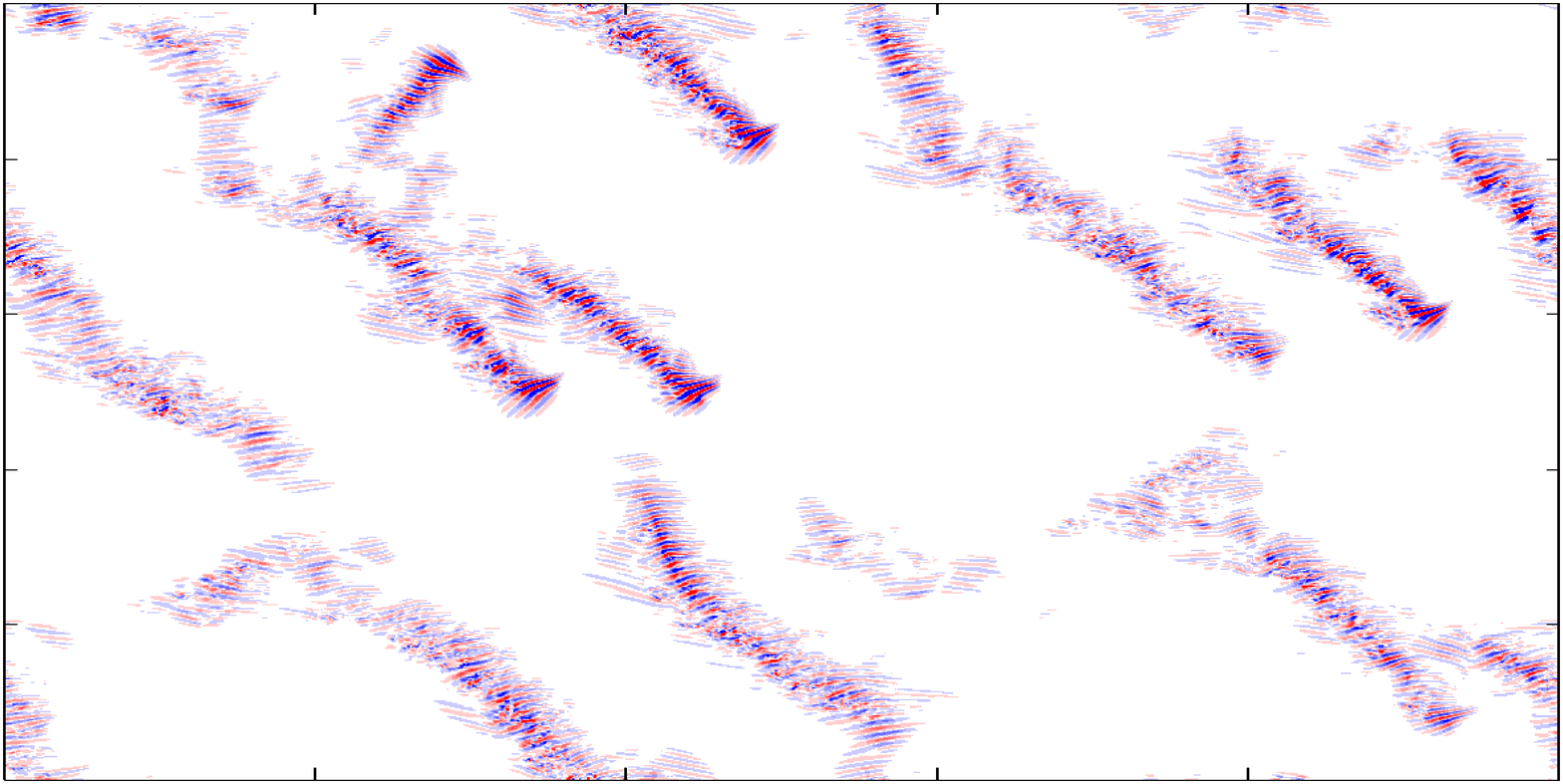}
\end{minipage}
\caption{ 
(a) Longitudinal splitting. 
(b) Longitudinal collision. 
(c) Transversal splitting.  
(d) Transversal collision. 
The same quantity as in Fig.~\ref{snap}, $u_y(x,0,z;t)$, is displayed in  a comoving frame with velocity $({u_x}^{\rm(f)},{u_z}^{\rm(f)})=(0.8,0.1)$.
\rev{Circles locate DAHs to be followed and numbers identify bands,
`P' and `C' stand for parent and child, respectively. }
See \rev{also} videos in~\cite{SM}.}
\label{4event}
\end{figure*}

Laminar--turbulent patterns below $\R=1200$ were examined 
to better understand the symmetry-restoring bifurcation observed at increasing $\R$. 
Processes involved in the dynamics are illustrated in Fig.~\ref{4event}.
The local spread and decay of turbulence respectively stem from splittings and 
collisions of LTBs with either identical or opposite orientations. 
Figure~\ref{4event}(a) shows the nucleation of a new band by 
{\it longitudinal splitting\/} of an LTB at its tail.
The active region is downstream and the splitting takes place upstream where the turbulence level is always weaker than near the DAH, contrary to what happens for puffs in pipe flow~\cite{A11,S14}.
When two parallel LTBs collide ({\it longitudinal collision}), following the large scale flow around them~\cite{X15,T15,T18}, the upstream faster LTB catches the downstream LTB that disappears as in Fig.~\ref{4event}(b).
At larger $\R$, another splitting process here called {\it transversal\/} can take place along an LTB:
A turbulent `bud' appears on the side of an LTB and forms an off-aligned turbulent branch as in Fig.~\ref{4event}(c).
Finally, when the DAH of a new-born LTB collides an LTB with a different orientation ({\it transversal collision}), the attacker most often dies in the collision, Fig.~\ref{4event}(d). 
The occurrence of transversal splitting causes the spread of turbulence to be a genuinely two-dimensional process.
Transversal splitting has also been observed in plane Couette flow and considered essential to the development of laminar--turbulent patterns~\cite{M12}.
Although transversal splittings are observed for $\R\ge 800$, one propagation direction remains dominant up to threshold $\R_2$.

This one-sided/two-sided spanwise-symmetry restoring bifurcation can be understood using a simple prey-predator model for the densities $X_\pm$ of two species of LTBs, left-propagating and right-propagating: 
\begin{gather}
\label{e1m}
{\rm d} X_+/{\rm d} t= a X_+ -bX_+^2 + c X_- - d X_+ X_-\,, \\
\label{e2m}
{\rm d} X_-/{\rm d} t= a X_- -bX_-^2 + c X_+ - d X_+ X_-\,.
\end{gather}
By construction, these equations incorporate the built-in spanwise symmetry of the system and each term corresponds to a process in Fig.~\rev{\ref{4event}}.
Coefficient $a$ represents the {\it longitudinal} splitting rate (Fig.~\rev{\ref{4event}}(a)).
The {\it transversal splitting\/} rate $c$ (Fig.~\rev{\ref{4event}}(c)), the natural control parameter, is assumed to increase with $\R$ according to the observations \rev{($c>0$ for creation of $X_\pm$ out of $X_\mp \ne0$, hence $c\ge0$ for $\R\gtrsim 800$, event A in Fig.~\ref{sketch})}.
Coefficients $b$ and $d$ account for the turbulence-level decrease by the collision between LTBs of either the same (Fig.~\rev{\ref{4event}}(b)) 
or different (Fig.~\rev{\ref{4event}}(d)) orientations.
For collisions between differently oriented LTBs, the term $- d X_+ X_-$ models the decay rate of one of the species $X_\pm$ taken as proportional to the cross-section of LTBs of the opposite kind $X_\mp$. 
Coefficient $d$, which is weakly dependent on $\R$  and a function of the speed of colliding LTBs, is assumed constant, as well as $b$ parameterizing a logistic self-interaction for predation among LTBs with the same orientation, i.e. $-bX_\pm^2$.
A reduced cross-section and a very small relative velocity between LTBs of the same kind suggest $b \ll d$.
\rev{By} contrast with works elaborating on reaction-diffusion models devised to account for local interactions in transitional flows \cite{barkley2016theoretical,shih2016ecological}, our approach is Landau-like and deals with global observables minimally coupled by purely phenomenological coefficients. 

The analysis of the model is straightforward when considering the total amount of turbulence $S=X_++X_-$ and the degree of asymmetry $A=X_+ - X_-$ as working variables. 
The equations for these variables become:
\begin{gather}
\label{e1m'}
\frac{{\rm d} S}{{\rm d} t}= 
(a+c) S - \mbox{$\frac12$} (b+d) S^2 - \mbox{$\frac12$} (b-d) A^2\,,\\
\label{e2m'}
\frac{{\rm d} A}{{\rm d} t}= (a-c) A - b S A\,.
\end{gather}
The two-sided regime labeled ``$**$'' corresponds to $A=0$, while $A\ne0$ implies the dominance of one propagation direction.
``$A_{**}=0$'' solves (\ref{e2m'}) in all circumstances.
Using (\ref{e1m'}), the symmetrical fixed point is then given by:
\begin{equation}
\label{fp**}
S_{**}=\frac{2(a+c)}{b+d}\,,\qquad A_{**}=0.
\end{equation} 
This fixed point has eigenvalues ($s_S,s_A$) with $s_S=-(a+c)<0$ and $s_A=[a(d-b)-c(d+3b)]/(b+d)$.
The symmetric solution is then stable as long as $s_A<0$; 
hence, $c>c_{\rm c}=a(d-b)/(d+3b)$ when $b<d$ as assumed from the observations.
The two-sided regime is then stable for large $\R$ and becomes unstable below a threshold corresponding to $c_{\rm c}$.

\begin{figure}
\includegraphics[width=\textwidth]{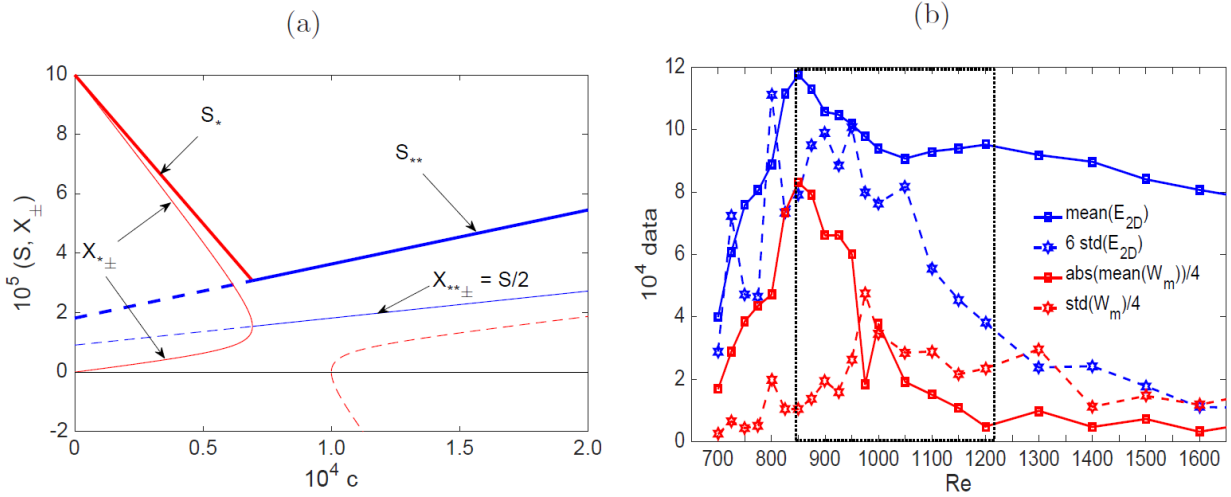}
\caption{\label{f-model}
Transition from one-sided to two-sided flow. (a) Model.
(b) Numerical simulation.}
\end{figure}

The one-sided regime labeled ``$*$'' corresponds to $A\ne 0$ at steady state (fixed point);\textbf{}
hence, fr\textbf{o}m~(\ref{e2m'}) and next from~(\ref{e1m'}):
\begin{equation}
\label{fp*}
S_{*}=\frac{a-c}{b}\,,
\quad{A_{*}}^2=\frac{(a-c)(d+3b)(c_{\rm c}-c)}{b^2(d-b)}\,.
\end{equation}
This analysis shows that the system experiences a standard super-critical pitchfork bifurcation toward asymmetry by decreasing $c$.
At $c=c_{\rm c}$ as previously defined, $(S_{**},A_{**})$ becomes unstable and is replaced by $(S_*,\pm A_*)$, which is stable for $c < c_{\rm c}$ as expected.
Figure~\ref{f-model}(a) displays the bifurcation diagram of model (\ref{e1m'},\ref{e2m'}) with splitting rate $c$ taken as the control parameter, and $a=10^{-4}$, $d=10$, $b=d/10$.
The thick line represents the total turbulence amount $S$ given by $S_{*}=(a-c)/b$ 
below $c_{\rm c}$ and $S_{**}=2(a+c)/(b+d)$ above.
Thin lines correspond to $X_\pm$.
\rev{Dashed} lines correspond to the unstable solutions (the \rev{dashed} 
branch for $c>a$ is furthermore irrelevant to the present problem since it leads to negative \rev{values of $X_{-}$}).

Experimental support to the model is displayed in the boxed region in Fig.~\ref{f-model}(b).
The total transverse perturbation energy $E_y+E_z=E_{2D}$ is taken as a proxy for $S$.
The mean spanwise velocity component $\Wm={\mathcal V}^{-1} \int_{\mathcal V} u_z {\mathrm d}{\mathcal V}$ is interpreted as an instantaneous measure of the degree of asymmetry $A$ because it cancels out statistically when symmetry is restored at high $\R$, i.e. $\average{\Wm}=0$, where $\average{\cdot}$ represents time average.
Notice that the standard deviation of $E_{2D}$ is multiplied by 6 and both $\average{\Wm}$ and its standard deviation are divided  by 4 so that variations of the observables can be more easily compared.
As long as transversal splitting is negligible, $\langle E_{2D}\rangle$ and $\average{\Wm}$ increase with the mean length of LTBs.
\rev{The two orientations correspond to} $\average{\Wm}$ with opposite signs.
Periodic spanwise boundary conditions leave open the possibility of having $\Wm\ne0$ as a result of symmetry breaking.
The case of an experimental system with solid lateral boundaries forbidding such a net transverse flux is considered later in \S\ref{S5}.
In Figure~\ref{f-model}(b) we only display $|\average{\Wm}|$ since at steady state, depending on the orientation of the LTBs, $\pm|\average{\Wm}|$ are equally possible.
Both $\langle E_{2D}\rangle$ and $|\average{\Wm}|$ reach their maximum values for $\R\approx850$, which can be understood as when transversal splitting --not observed below $\R=800$-- becomes significant.
For $\R>850$, our two observables follow the trend suggested by the model:
$\langle E_{2D}\rangle$ decreases roughly linearly as $\R$ increases up to $\R \approx 1000$ and slowly grows beyond, in agreement with equations (\ref{fp**}, \ref{fp*}).
Likewise, $|\average{\Wm}|$ decreases rapidly to zero, similar to $A_*$, as indicated by the fact that the standard deviation becomes larger than the mean, at $\R\approx 1000$.
For larger $\R$ outside the box, the system enters a developed two-sided regime where the model, designed to account for the one-sided/two-sided bifurcation, becomes insufficient.
Inside the box, this oversimplified formulation well captures the phenomenology of the transition at a qualitative level.

\begin{figure}
\begin{center}
\includegraphics[width=0.5\textwidth]{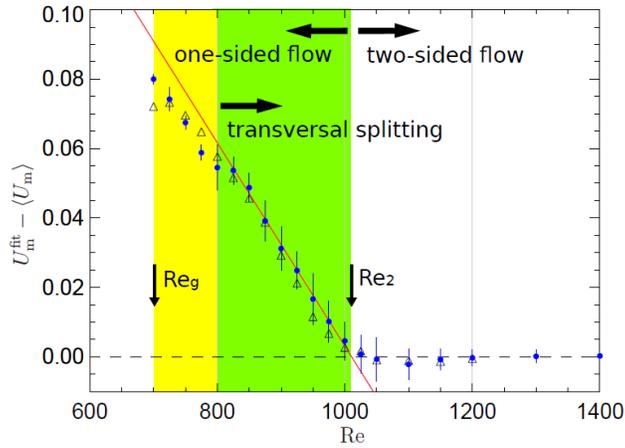}
\end{center}
\caption{Departure of observed streamwise mean flow $\average{\Um}$ from the value predicted by the law $\Um^{\rm fit}= 19.47/\sqrt{\R}$, solid line in Fig.~\ref{mean2}(a).  
Fitting the data against a straight line for $850 \le \R \le 1000$ yields $\R_{2}\simeq1011$. 
Triangles are for the larger domain $1000\times2\times500$.}
\label{ft}
\end{figure}

The transition is then understood by assuming that parameter $c$ measuring the transversal splitting rate increases with $\R$.
The analysis above shows that $c_{\rm c}\simeq a$  assuming that $b \ll d$, as  stems from our observations. 
Furthermore, in the range $850 \le \R \le 1200$, the variations of $S$ and $A$ around $c_{\rm c}$ are consistent with those of the turbulent energy and the spanwise mean velocity, respectively.
In fact, the deviation of $\average{\Um}$ away from its behavior in the two-sided regime fitted as $\Um^{\rm fit}=w/\sqrt{\R}$ in Fig.~\ref{mean2}(a) is a \rev{proxy}  for the change of total amount of turbulence at the one-sided/two-sided bifurcation.
As displayed in Fig.~\ref{ft}, this variation is linear close to the bifurcation point $\R_2\simeq1011$, in agreement with the predictions of the model. 
As can be seen in the inset of Fig.~\ref{mean2}(a)  and in Fig.~\ref{ft}, the current domain size is appropriate to obtain this threshold.

\section{Behavior above $\R_2$\label{S4}}

\begin{figure}
\begin{minipage}{0.49\textwidth}
\begin{center}
$u_x(y=0)$ \\[0.5ex]
\includegraphics[width=0.96\textwidth]{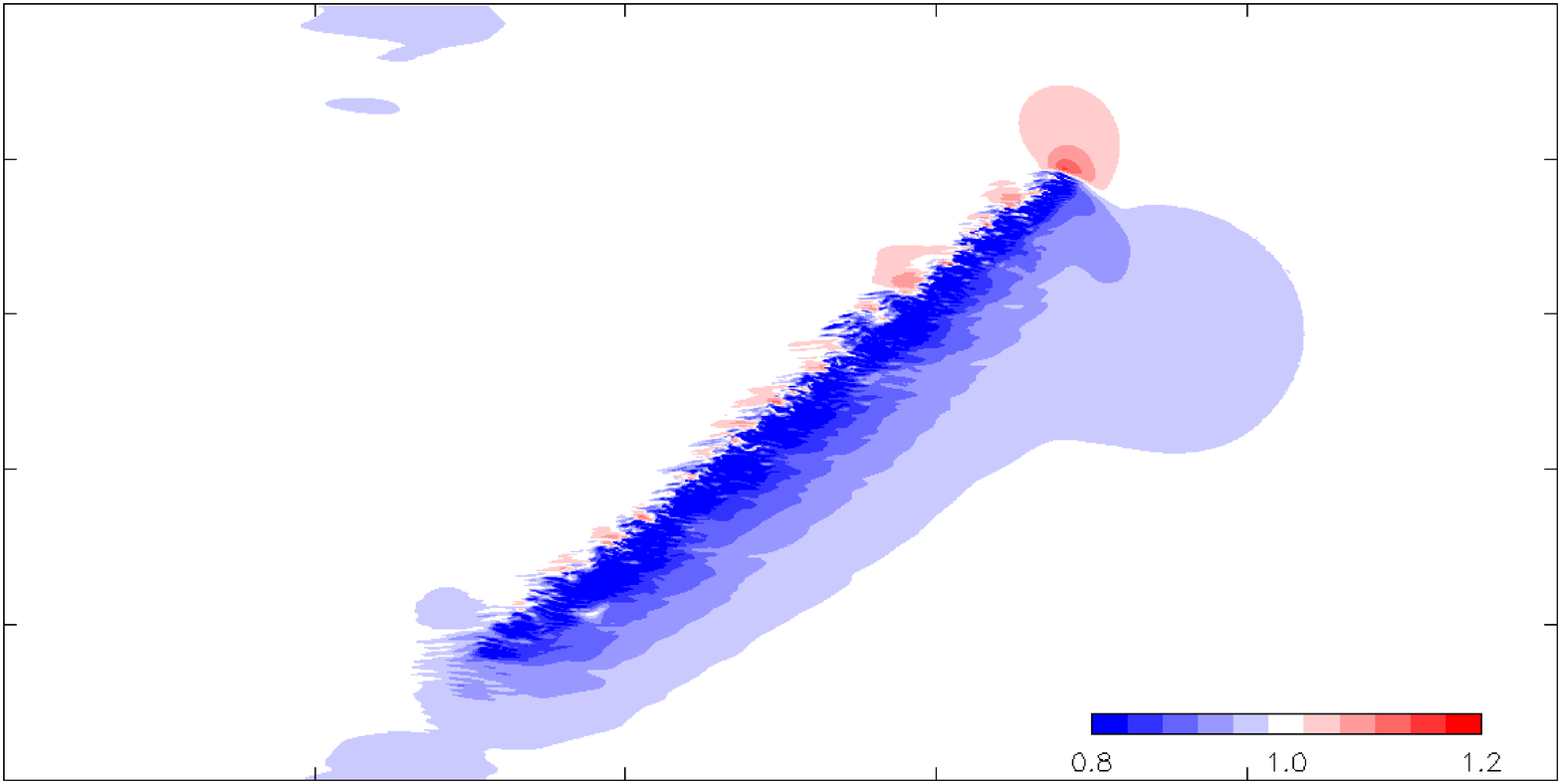}
$u_z(y=0)$ \\[0.5ex]
\includegraphics[width=0.96\textwidth]{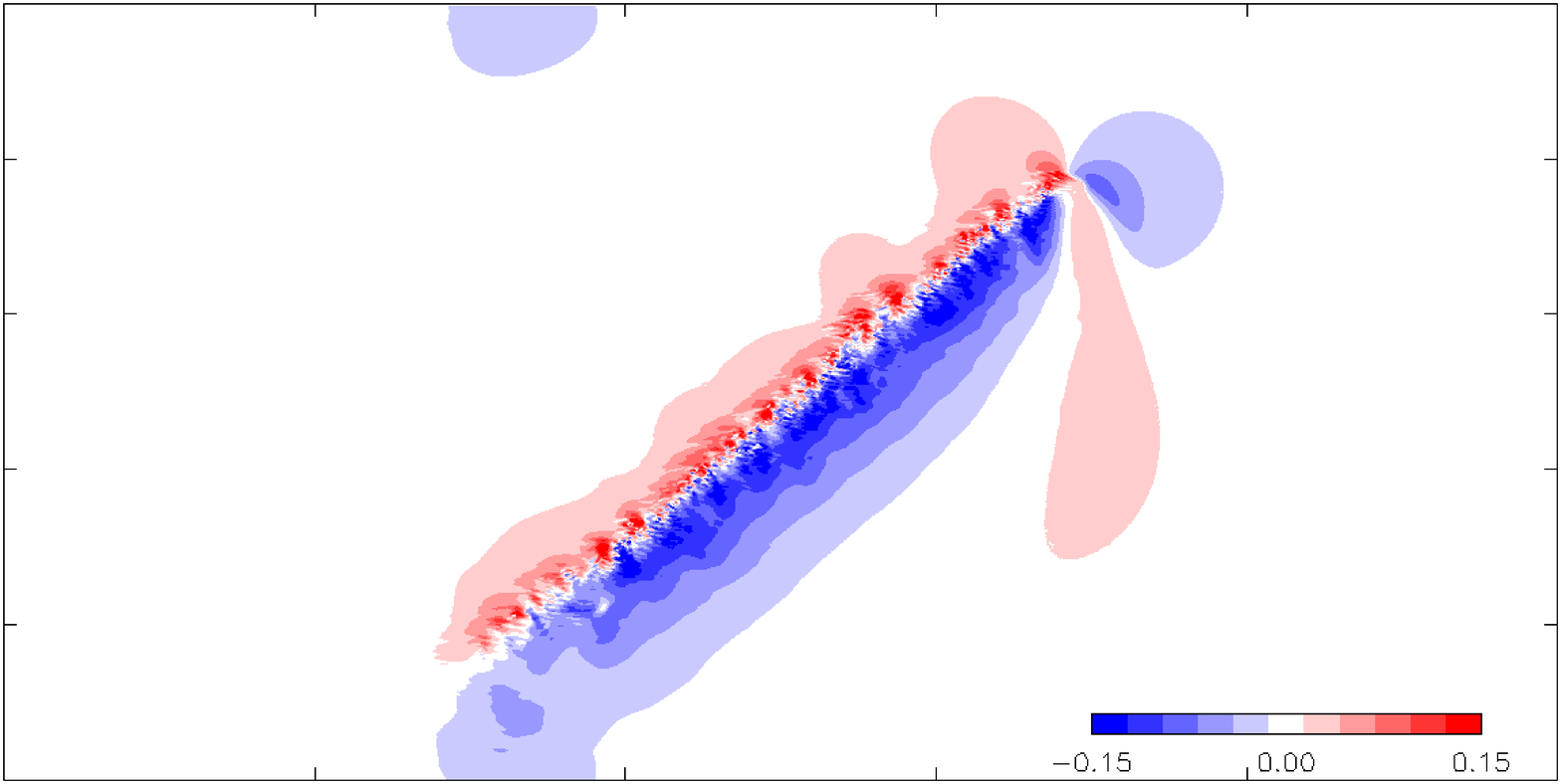}
\end{center}
\end{minipage}			
\begin{minipage}{0.49\textwidth}
\begin{center}
$u_y(y=0)$ \\[0.5ex]
\includegraphics[width=0.96\textwidth]{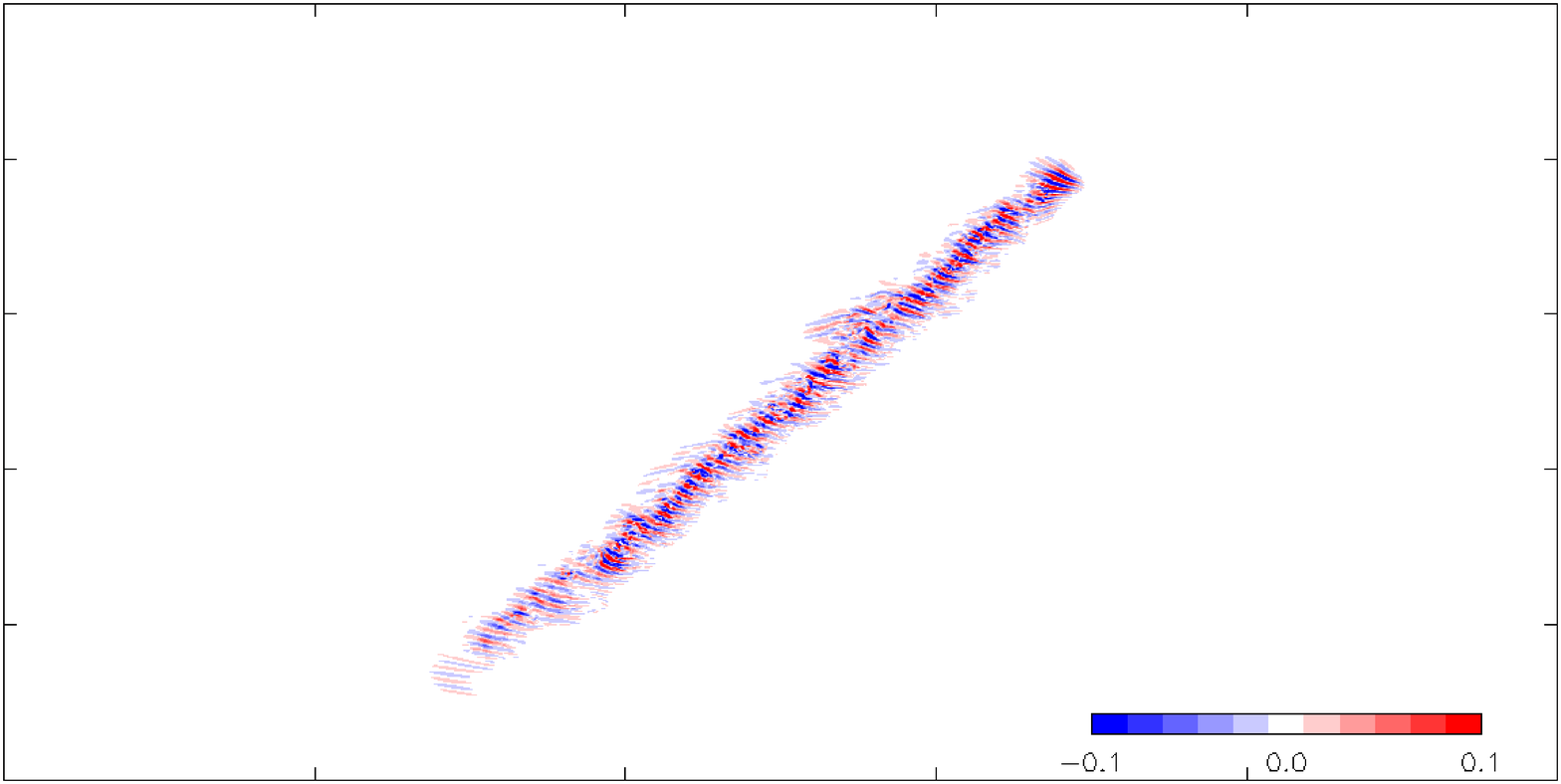}\\
$\average{u_y^2+u_z^2}_y$ \\[0.5ex]
\includegraphics[width=0.96\textwidth]{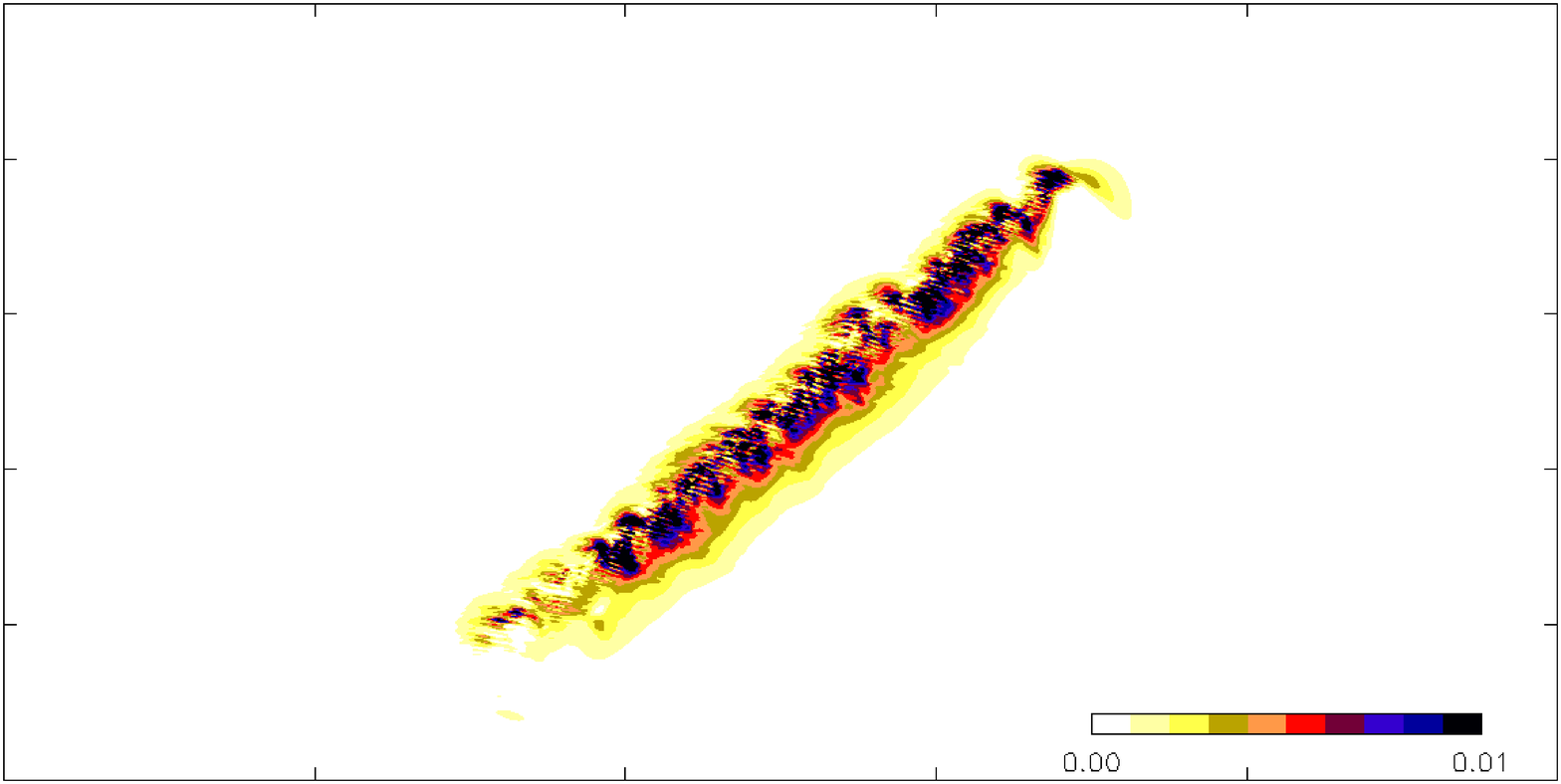}
\end{center}
\end{minipage}
\caption{Possible candidates for the determination of the turbulent fraction illustrated by ${2D}$-images of the same localized turbulent band at $\R=700$. The flow is from left to right and the downstream active head is in the upper right corner in each panel. \rev{Domain size is $500\times250$}.}
\label{Sltb}
\end{figure}

Beyond $\R_2$ and up to entrance in the tight-banded pattern regime, Fig.~\ref{snap}(e), channel flow exhibits a spatiotemporal intermittent behavior, Fig.~\ref{snap}(c,d), strongly reminiscent of DP above the threshold. 
The work of Sano and Tamai \cite{ST16} focusing on the critical properties of turbulence decay in a DP context \cite{H08} naturally suggests one to study this Reynolds number range in terms of turbulent fraction $F_{\rm t}$.
This in turn relies on the identification of appropriate laminar and turbulent local states, based on observables that vary sufficiently sharply in space to define the respective domains properly, allowing a precise measurement of their relative occupancy fractions.
\revv{Figure} \ref{Sltb} displays four possible candidates, the three velocity components in the mid-gap plane $y=0$, and the mean transverse perturbation energy $E_{2D}(x,z;t)=\frac12\int_{-1}^{1} (u_y^2+u_z^2)\, {\rm d}y$ around \rev{an} LTB at $\R=700$.
In-plane components $u_x$ and $u_z$ display a large-scale, slowly decaying, structure around the LTB.
\rev{By} contrast, $u_y$ sharply discriminates non-laminar flow regions.
$E_{2D}$ is slightly less close-fitting due to the limited  contribution of the component $u_z$.
Accordingly, the absolute value of the wall-normal velocity on the mid-plane $|u_y(x,0,z)|$ will be used to evaluate $F_{\rm t}$.
However, inside LTBs, $u_y$ displays small-scale oscillations associated with the presence of streamwise vortices, which produces narrow regions  with $|u_y(x,0,z)|\approx0$ that must not be counted as laminar.
Accordingly, the field $|u_y(x,0,z)|$ has to be smoothed  beforehand, here by simple box-averaging over cells of size $w\times w$, and next thresholded using the ``moment-preserving'' procedure \cite{T85}, as explained in Appendix~\ref{AppB}.

\begin{figure}
\includegraphics[width=\textwidth]{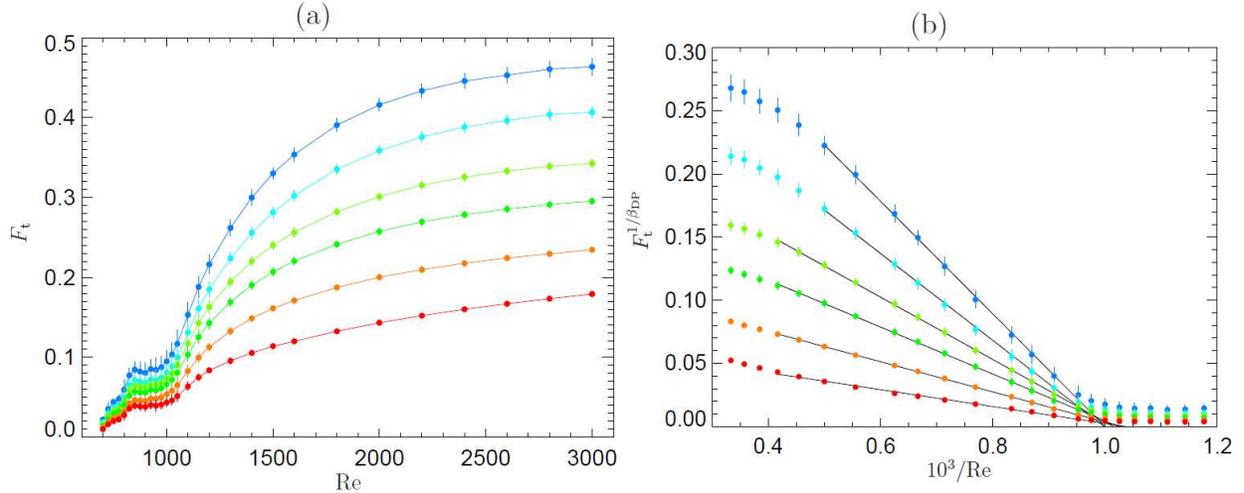}
\caption{$F_t$ for different $w$. 
From the bottom to the top: $w=0$ (no filtering, red), 
$4\delta$, $8\delta$, $12\delta$, $20\delta$, and $40\delta$ (blue). 
(a) $F_{\rm t}$ as a function of $\R$.
(b) $F_t^{1/\beta_{\rm DP}}$ as a function 
of $1/\R$  with $\beta_{\rm DP}=0.583$.
\label{Sft}}
\end{figure}

The variation of $F_{\rm t}$ as a function of $\R$ is displayed in Fig.~\ref{Sft}(a) for $700\le \R \le 3000$ and  for $w$ varying between~0 (no filtering) and $40\delta$.
For all values of $w$, similar variations are observed and one readily identifies the different stages illustrated in Fig.~\ref{snap}:
\rev{one-sided growth for Re $\le \R_2 \simeq 1011$, with a clear change of regime for $\R \sim 850$ as transversal splitting sets in,}
before a rapid increase akin to a power-law growth as $\R$ increases in the symmetry-restored \revv{regimes} 
(\revv{Fig.~\ref{snap}(b)-(d)}) up to the tight banded pattern regime 
(\revv{Fig.~\ref{snap}(e)}). 

The consistent square-root-like behavior is reminiscent of the growth of the order parameter of DP beyond threshold.
In line with the idea that the critical properties of the stochastic DP process are relevant for channel flow, as put forward by Sano \& Tamai, we first test the plausibility of exponent $\beta_{\rm DP}\simeq0.583$ to describe the variation of the turbulent fraction with the relative distance to threshold.
\revv{Figure}~\ref{Sft}(b) therefore displays $F_{\rm t}^{1/\beta_{\rm DP}}$ as a function of $1/\R$.
The linear behavior of the plots for different values of $w$, systematically extrapolating to zero for around $\R\approx 1000$ ($1000/\R\approx1)$ therefore strongly supports the expected behavior of $F_{\rm t}$ as if it were produced by a $2D$-DP process. 
From the least-square fits, the straight lines in \revv{Fig.}~\ref{Sft}(b), it is seen that some filtering is needed to eliminate spurious small scale oscillations of $u_y(x,0,z)$ since raw data ($w=0$, red data points) does not behave so satisfactorily.
On the other hand, strong filtering ($w\ge 20$) leads to a reduction of the range where a good linear fit is obtained ($1050\le \R \le 2400$ for $w \le 12\delta$, and $1050\le \R\le 2000$ for $w=20\delta$ and $40\delta$).
To further develop the quantitative analysis we choose \rev{the} best compromise which seems to be $w=12\delta$, i.e. the largest filtering possible in the \revv{widest} 
$\R$ interval with the expected property. 

\begin{figure}
\includegraphics[width=\textwidth]{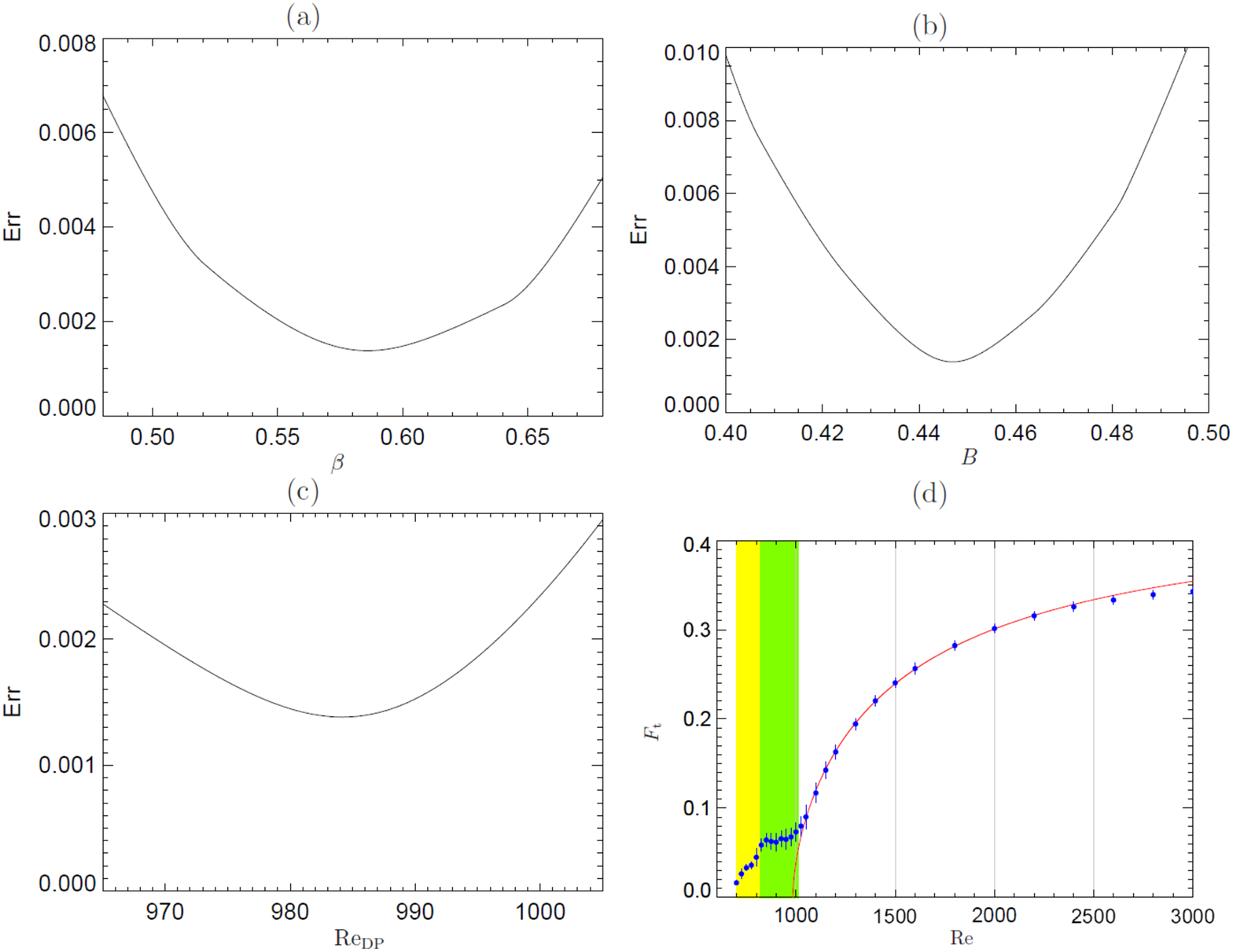}\\
\caption{\label{Serr}
(a--c) Minimum of ${\rm Err}$ as a function of the different fit parameters for $\Ft$ with $w=12\delta$ and control parameter $1/\R$.  
(a) Exponent $\beta$. (b) Amplitude $B$. 
(c) Threshold $\R_{\rm DP}$.
\rev{In these panels, the error is displayed as a function of one fitting parameter, the two other being supposed fixed at their optimal value.}   
(d) Turbulent fraction $\Ft$ as a function of $\R$ and its fit against the theoretical expression given in \S\ref{S4}
}
\end{figure}

In these conditions, a direct fitting over the interval $1050 \le \R \le 2400$ has been attempted against the function $\Ft = B \left(1-\R_{\rm DP}/\R \right)^\beta$ with $B$, $\R_{\rm DP}$ and $\beta$ as fitting parameters.
We proceeded to a least-square minimization of the  error ${\rm Err}^2=$\\
\noindent
$\frac{1}{N}\sum_{n=1}^{N}  \left( {\Ft}_n-B \big(1-\R_{\rm DP}/\R_n \right)^\beta \big)^2$ where $N$ is the number of values $\R_n$ of $\R$ entering the fit, and ${\Ft}_n$ the corresponding measured mean turbulent fraction. 
\revv{Figure}~\ref{Serr}(a) displays the minimum error as a function of $\beta$, pointing to $\beta\approx0.58$, while panels (b) and (c) display similar results for $B\approx0.45$ and $\R_{\rm DP}\approx984$ as best fitting values, respectively.
Following this approach, the estimate for $\beta$ turns out to be close to the theoretical value $\beta_{\rm DP}=0.583$ that served as initial guess in Fig.~\ref{Sft}(b).
The error curves in each case however indicate that the optima are not sharply defined.
Unfortunately, these estimates cannot be improved mainly because the critical point cannot be approached sufficiently closely.
This also justifies that we did not try to adjust the other critical parameters related to space-time correlations since they are discriminating solely in an arbitrarily close neighborhood of $\R_{\rm DP}$.

Our results, used to draw the fitting line in Fig.~\ref{Serr}(d), therefore point toward a plausible universal DP behavior for large enough turbulent fraction, a behavior inevitably truncated before criticality by a crossover to a different decay regime as $\R$ decreases.
As a matter of fact, the turbulent fraction becomes so small that laminar gaps open along oblique arms of the fluctuating laminar-turbulent network, which produces the development of DAHs --original to channel flow-- so that the DP regime is superseded by the LTB-dominated regime and, as $\R$ decreases even more, to the symmetry-breaking bifurcation below which turbulence decay takes up a novel turn.

\section{Discussion and concluding remarks\label{S5}}

Apart from a few illustrations of the laminar-turbulent patterning in the upper transitional range of channel flow in Fig.~\ref{snap}(e,f), our results all relate to its lower part featuring the decay to laminar flow where a DP scenario with universal properties expected for two-dimensional systems has been put forward~\cite{ST16}, while sustained LTBs have been observed at values of $\R$ lower than the measured DP critical point~\cite{Tao13,X15,T15,K17,T18,H16,Pa17}.
The first result comes in support of a theoretical conjecture by Pomeau~\cite{P86} based on the recognition that the flow can be locally in one of the two possible states, turbulent or laminar.
Furthermore, in the transitional range, developing turbulence presents itself as a contamination of the laminar state, locally linearly stable and therefore {\it absorbing\/}, by a highly fluctuating, unstable, chaotic {\it active\/} state.
This framework is directly derived from a statistical-physics approach where the key concept is the thermodynamic limit of infinitely wide systems at statistically steady state (here non-equilibrium steady state).
Though experiments and simulations are performed in domains and for durations somewhat far from the thermodynamic-limit conditions (with~\cite{C17} as a possible exception), the question of what we may infer on general grounds from our results is relevant.

Let us first notice that comparisons with laboratory experiments and computer simulations are made difficult owing to the several possibilities chosen to drive the flow, either generated by a fixed pressure gradient (or given average friction at the walls) or a constant mean streamwise flow rate.
Though it is generally admitted that the results should be statistically identical at the thermodynamic limit, quantitative correspondence between different works is not easy to establish, and first of all with respect to how $\R$ is defined owing to what is controlled and what is measured.
Here we fix the mean pressure gradient (fixed bulk force $f$) and use the computed centerline velocity of the corresponding laminar flow $U$ as speed unit, while the half-distance $h$ between the plates is taken as length unit and $h/U$ as time unit.
The definition of $\R=Uh/\nu$ follows.
If wall units are preferred, e.g. in~\cite{T05}, one gets $\R_\tau = \sqrt{2\R}$ as derived in \S\ref{AppA}.
In the laminar regime the mean flow rate is $\revv{\Um}=\frac23 U$ and the Reynolds number made out of it: $\R_{\rm m}=\revv{\Um} h/\nu = \frac23 \R$ is frequently used.
However, when the flow is not fully laminar, the measured mean flow rate  $\langle U_{\rm m}\rangle$ decreases below its nominal value $2/3$ (Fig.~\ref{ft}).
Accordingly, the Reynolds number constructed using $\langle U_{\rm m}\rangle$ is not a control parameter but an observable that has to be determined by averaging $u_x$ over space and time.
This lead us to define $\Rm = \frac32 \average{\R_{\rm m}}=\frac32 \average{U_{\rm m}} h/\nu$.
With this definition, when the flow is fully laminar, $\Rm\equiv\R$ and, accordingly, the linear instability threshold is $\R_{\rm TS}=5772$~\cite{Or71}.
Willing to keep $\langle U_{\rm m}\rangle$ strictly constant, as done in some numerical experiments and only approximately achieved in the laboratory, is therefore a different experiment in which the applied pressure gradient is unknown and fluctuating. 
All in all, our results in Fig.~\ref{snap} are consistent with those found in the literature when expressed in terms of $\Rm$, especially those in~\cite{T05,T18}, or even in~\cite{Tu14,NB}.

Our main contention is that the previously mentioned conflicting experimental observations about DP universality and the existence of LTBs can be reconciled provided that it is recognized that the DP scenario cannot be followed down to its critical point.
We have shown that, when the turbulent fraction is small enough, processes develop which are not present in other systems where DP-universality is observed:
As $\R$ is slowly decreased, the flow evolves from the weakly intermittent loose banded pattern regime, Fig.~\ref{snap}(d), to the strongly intermittent continuous network regime, Fig.~\ref{snap}(c).
Further decreasing $\R$ induces the opening of laminar gaps along the turbulent branches, here called {\it splittings\/}, and the formation of LTBs terminated by DAHs constitutive of the two-sided LTB regime, Fig.~\ref{snap}(b).
At first the intermittent expansion/recession of turbulence is two-dimensional in the plane of the flow owing to comparable rates for transversal and longitudinal splittings, but the rate of transversal splittings decreases with $\R$ faster than that of longitudinal splittings, which implies a change to one-sided LTB propagation, i.e. a symmetry breaking due to a deficit of regeneration of one of the two LTB species at a local scale.
In the long term, the rate difference indeed brings about a global dominance of one orientation over the other and a mostly one-dimensional expansion/recession of turbulence.
In this respect the cases of Waleffe flow~\cite{C17} \revv{and} 
plane Couette flow~\cite{Sh17} appear different.
In these two cases, laminar gaps also open through splittings, longitudinal and transversal, along the turbulent segments forming the fluctuating laminar-turbulent network \rev{but  the upstream-downstream distinction does not make sense.
Close to $\Rg$, the turbulent patches accordingly take the form of fluctuating short oblique straight segments, v-shaped, or x-shaped spots~\cite[Fig.~2]{C17} with dynamics profoundly different from that of LTBs equipped with DAHs. 
As a result the turbulent fraction can then decrease indefinitely according to $2D$-DP universality, down to $\R_{\rm DP}$ that is reached before any symmetry-breaking bifurcation has a chance to take place, presumably because the longitudinal and transversal splitting rates remain comparable as $\R$ decreases.}
\rev{By} contrast, specificities of channel flow prevent the observation of a complete $2D$-DP scenario and the origin of this {\it imperfection\/} is fully characterized as the result of an unavoidable symmetry-breaking bifurcation when the turbulent fraction becomes small enough, before the expected $\R_{\rm DP}$ is reached.

We now focus the discussion on two main points: the variation of the turbulent fraction in the range $1050\le \R\le 2400$ in connection with the DP scenario above the symmetry-breaking bifurcation, and the non-zero mean transverse flow in the symmetry broken regime, $\R\le \R_2$, its observability at the thermodynamic limit or, at the opposite, in laterally bounded finite-width channels.

For the conceptual reasons already evoked, DP is relevant  {\it as a process\/} in the lowest transitional range.
In its simplest formulation used to typify the universality class, a single contamination probability is needed to control the transition but the nature of physical mechanisms involved and their dependence on the Reynolds number are not known.
The expectation for universality rests on the Janssen--Grassberger conjecture~\cite{H08} stipulating in particular that the absorbing state is unique, the interactions are local, and that there are no weird parasitic effects such as quenched disorder, which is implicitly assumed if Pomeau's educated guess is valid.
Strictly speaking, the question whether the decay of turbulence in channel flow follows $2D$-DP universality is void since the critical regime -- the immediate neighborhood of the threshold -- cannot be entered, but results presented in~\S\ref{S4} suggest a positive answer if we accept to loosen this restriction (Fig.~\ref{Serr}).

This argument however hides a difficulty since, as already mentioned, the relation between the chosen external control parameter and the internal control parameter, the effective contamination probability generated by stochastic processes at the scale of a few MFUs, is not known.
On general grounds, the critical behavior is not sensitive to the choice of the external control parameter in an asymptotically close vicinity of the critical point while departures from universality become conspicuous at variable distances from the threshold depending on that choice.
In the previously reported cases, the variation of the turbulent fraction as a function of the Reynolds number deviates from the expected power law dependence beyond a relative distance $\epsilon$ to the critical point  $\sim 10^{-2}$ for quasi-$1D$ plane Couette flow~\cite[Fig.~2]{L16}, or $\sim 10^{-3}$ for quasi-$2D$ Waleffe flow~\cite[Fig.~4]{C17}.
For channel flow, the universal behavior of the turbulent fraction is reported for $\epsilon \lesssim 0.1$ in \cite[Fig.~3]{ST16}, but the experimental conditions do not permit access to the LTB regime.
\rev{By} contrast, taking advantage of a streamwise periodically continued domain, our approach leaves sufficient time for turbulent patches to rearrange into LTBs, which truncates the DP behavior above threshold~\cite{NRm}.
Our most striking finding is that, characterizing the flow by its natural control parameter $\R$, the turbulent fraction then varies in accordance with universality \rev{over a} surprisingly wide interval from below $\R\approx 2400$ nearly down to $\R_2$ where the dynamics becomes controlled by LTBs.
When expressed in terms of $\epsilon=1-\R_{\rm DP}/\R$ with $\R_{\rm DP}=984$ as the extrapolated threshold, this range extends over $0.07\lesssim \epsilon\lesssim0.6$ (Figs.~\ref{Sft}(b) and \ref{Serr}(d)).
At this point, it is fair to add that other possibilities for the external control, namely using $\R_\tau$ or $\Rm$ in principle equivalent in the neighborhood of the threshold, perform badly at some distance above threshold, not even permitting to detect a neat power-law variation over such a wide interval.
The physical meaning of $\R_\tau$ or $\Rm$ implies an averaging over the flow in a strongly inhomogeneous spatiotemporally intermittent state, which might explain that the probabilities issued from the dynamics at the MFU scale are less straightforwardly parameterized using them than using $\R$ (or $1/\R$).

Let us now turn to the symmetry breaking observed at decreasing $\R$ and studied in \S\ref{S3}.
Elementary events involved in the dynamics of LTBs were described in detail (Fig.~\ref{4event}).
A satisfactory simple phenomenological model was developed then in the spirit of Landau theory to account for the bifurcation at a qualitative and semi-quantitative level.
Extrapolating our results in a finite size domain with periodic boundary conditions to a laterally unbounded system (thermodynamic limit), when transversal splitting is negligible ($\R\lesssim800$) one-sided propagation is expected because for $\R$ low enough, any transversal collision destroys one of the colliders (Fig.~\ref{4event}(d)) and, as seen in the movies at low $\R$~\cite{SM}, the transient evolution of sparse turbulence follows a majority rule.
At short times, a patchwork of domains with one or the other dominant orientation forms, with transversal collisions concentrated along the domain boundaries.
At longer times, a ripening process develop during which domains compete with each other, locally applying the majority rule so that, 
in the very long time limit at very low LTB density, one can figure out an asymptotic steady state which is mostly uniformly oriented.
Residual LTBs are then moving nearly at the same speed, along the same direction, and interact only through longitudinal splittings/collisions (Fig.~\ref{4event}(a,b)) maintaining a state free of transversal collisions.
Such a regime is expected to persist down to $\Rg$ below which it decays, much like in pipe flow~\cite{A11}, due to the predominance of longitudinal destructive collisions over expanding turbulence through splittings to be discussed in a future publication~\cite{SMxx}.

The difficulty with this picture is that, on general grounds, such a symmetry breaking is associated to a large scale flow with a transverse component corresponding to a net, fluctuating, spanwise mean flux with non-zero average $\average{\Wm}$.
In systems with periodic spanwise boundary conditions, symmetries do not forbid the existence of such a flow.
Close to the bifurcation point, for $\R<\R_2$ when transversal splitting plays a significant role, the patchwork alluded to above  can reach a statistically steady state with $\average{\Wm}\ne0$ but reduced by compensations between patches of different orientations, whereas for $\R>\R_2$ one gets $\average{\Wm}\approx0$ with substantial fluctuations, see Fig.~\ref{f-model}(b).
These are the characteristics of an order parameter apt to quantify the symmetry-breaking bifurcation. 
From the observations reported in the previous paragraph, these properties are expected to hold also as $t$ goes to infinity in a spanwise unbounded system (thermodynamic limit).

In finite-width channels with impervious lateral walls, \rev{any net spanwise mean flow $\Wm$} is not permitted.
In laboratory experiments where this condition applies, the one-sided regime has been observed~\cite{H16,Pa17} but, in contrast with our simulations where LTBs drift obliquely, after an initial transient stage during which the flow equilibrates, they are advected  strictly along the streamwise direction.
This means that an additional spanwise pressure wave accompanies the passing of an LTB, producing a spanwise mean flow component able to deviate it, thus compensating for the \rev{$\Wm$} that it \rev{would} naturally  \rev{generate}.
Rigid lateral walls are obviously able to withstand such pressure fluctuations as LTBs pass by. 
Only simulations specially designed to implement the corresponding no-slip lateral boundary conditions, e.g. with spanwise Chebyshev polynomials~\cite{Tetal15}, would permit a detailed account of the LTB propagation in the one-sided regime.
However, while keeping spanwise periodic boundary conditions, one can think of correcting our governing equations for an additional spanwise fluctuating bulk force generating a mean transverse flow sufficient to compensate the drift of LTBs and statistically maintain a strictly streamwise propagation.
Such a work remains to be done but could give hints on the one-sided regime in realistic experimental conditions, owing to the obvious robustness of LTBs.

Two complementary studies are in progress, above and below the range of Reynolds numbers considered in this work: one is dedicated to the decay of the one-sided regime in a domain $1000\times2\times500$ or larger~\cite{SMxx} and the other is focussed on the onset of the laminar-turbulent patterning {\it via\/} standard Fourier analysis rather than turbulent fraction determination~\cite[see Fig.~4 for preliminary results]{MS19}.

By way of conclusion, the introduction of concepts and methods of statistical physics, and notably directed percolation, have put stress on universal features of the transition from/to turbulence in wall-bounded shear flows.
All over the spatiotemporally intermittent regimes along the transitional range of channel flow, laminar-turbulent coexistence with a high level of stochasticity legitimates Pomeau's views~\cite{P86} about the decay of turbulence at $\Rg$ as a process in the $2D$-DP universality class.
We have shown that this claim is however only partly fulfilled because a specific phenomenon comes and renders the full scenario imperfect:
spanwise symmetry is broken due to a sensitive balance between local processes (longitudinal {\it vs.} transversal splittings) with rates depending on $\R$ in different ways, ending with recession/expansion of turbulence becoming mostly one-dimensional before $2D$-DP criticality has a chance to be observed.
In turn this symmetry breaking is an event that could be well understood within the standard framework of dynamical systems and bifurcation theory, bringing an original perspective to the debated issue of universality {\it vs.\/} specificity in the transition to turbulence of wall-bounded flows.

\begin{acknowledgments}
We would like to thank Y. Duguet (LIMSI, Orsay, France) for interesting discussions about the problem.
\rev{Referees should also be thanked for their comments that helped us improve our work.}
Travel support is acknowledged from CNRS and JSPS through the collaborative project {\sc TransTurb}. 
This work was specifically supported by JSPS KAKENHI (JP17K14588), NIFS Collaboration Research program (NIFS16KNSS083), and Information Technology Center, The University of Tokyo. 
\end{acknowledgments}

\appendix

\section{System and numerical procedures}
\label{AppA}

We consider the flow between two parallel walls driven by a time-independent 
body force, usually called channel flow or plane Poiseuille flow.
The equations governing the velocity field $\bm u$ are as follows:
\begin{gather}
\label{e0}
\mbox{(a)} \quad \nabla \cdot \bm u =0\,,
\hspace{3em}\mbox{(b)}\quad
\frac{\partial \bm u}{\partial t} + (\bm u \cdot \nabla) \bm u = 
- \frac{1}{\rho} \nabla p  + \nu \Delta \bm u + f \hat{\bm e}_x\,,
\end{gather}
where $\rho={\rm Const.}$ is the density, $\nu$ the kinematic viscosity, and $f$ 
the body force specific density. 
The unit vector in the streamwise direction $x$ is denoted as $\hat{\bm e}_x$.
The $y$- and $z$-axes are along the wall-normal and spanwise directions, respectively.
All fields are assumed in-plane periodic and the velocity fulfills the usual no-slip 
boundary conditions at the walls.

The center-plane velocity of the corresponding laminar flow is $U=f h^2/2\nu$, 
where $2h$ is the distance between the walls. 
The Reynolds number is defined as $\R=U h/\nu = f h^3/2\nu^2$. 
Using $h$ and $U$ as distance and velocity units, and $h/U$ as the time unit, 
all variables below are assumed dimensionless without a 
notational change, the equations governing the velocity field $\bm{u}(x,y,z,t)$ 
are as follows: 
\begin{gather}
\label{e1}
\mbox{(a)} \quad \nabla \cdot \bm u =0\,,
\hspace{3em}\mbox{(b)}\quad
\frac{\partial \bm u}{\partial t} + (\bm u \cdot \nabla) \bm u = -\nabla p  +\frac{1}{\R} 
\Delta \bm u + \frac{2}{\R} \hat{\bm e}_x .
\end{gather}

In practice, these equations are rewritten for the wall-normal velocity component 
$u_y(x,y,z,t)$ and the wall-normal vorticity $\omega_y(x,y,z,t)=(\nabla \times {\bm u})_y$ 
obtained by applying  $\nabla \times$ and $\nabla \times \nabla \times$ to (\ref{e1}b) and 
keeping its $y$ component~\cite{SH01}:
\begin{gather}
\label{e2}
\mbox{(a)}\quad \frac{\partial \omega_y}{\partial t} = 
(\nabla \times {\bm N})_y + \frac{1}{\R} 
\Delta \omega_y\,,\hspace{3em}\mbox{(b)}\quad
 \frac{\partial \Delta u_y}{\partial t}   = 
-(\nabla \times  \nabla \times {\bm N})_y 
+ \frac{1}{\R} \Delta \Delta u_y\,,
\end{gather}
where ${\bm N}= \bm u \times \bm \omega$.

The full solution requires additional 
equations for in-plane averaged velocity fields.
We define the auxiliary fields $\phi_x(y,t)$ and $\phi_z(y,t)$ as 
$\partial_y \phi_x 
= \average{u_x}_{xz}$ and $\partial_y \phi_z = \average{u_z}_{xz}$, 
where
$\average{g}_{xz}$ denotes in-plane averaging of field $g$, 
namely $(L_x L_z)^{-1} \int_{0}^{L_z} \int_{0}^{L_x} g\, 
{\rm d}x\,{\rm d}z$.
By averaging the $z$ and $x$ components of the vorticity equation 
$\nabla \times$ (\ref{e1}b) we obtain:
\begin{gather}
\label{e3}
\mbox{(a)}\quad \frac{\partial}{\partial t}  \frac{\partial^2 \phi_x }{\partial y^2} 
 = \frac{\partial \average{{\bm N}_x}_{xz}}{\partial y} + \frac{1}{\R} 
 \frac{\partial^4 \phi_x }{\partial y^4}\,,
 \hspace{3em}\mbox{(b)}\quad
\frac{\partial}{\partial t}  \frac{\partial^2 \phi_z}{\partial y^2} 
 = \frac{\partial \average{{\bm N}_z}_{xz}}{\partial y} + \frac{1}{\R} 
 \frac{\partial^4 \phi_z}{\partial y^4}\,.
 \end{gather}
Fields $\phi_x$ and $\phi_z$ are defined up to arbitrary 
functions of time that can be fixed as follows: 
The streamwise bulk velocity is given by the difference between 
the boundary values of $\phi_x$ at $y=\pm1$, 
$\Um=\frac12 \int_{-1}^{+1} \average{u_x}_{xz}{\rm d}y
=\frac12[\phi_x(y,t) ]_{y=-1}^{y=+1}$.
The arbitrariness in the definition of $\phi_{x}$ can then be lifted 
by choosing $\phi_x(+1,t)= \Um(t)$; hence, $\phi_x(-1,t)= -\Um(t)$.
Similar conditions apply to $\phi_z$ and $\Wm$.

Equations for $\Um(t)$ and $\Wm(t)$ are obtained by 
averaging (\ref{e1}b) with respect to full space $(x,y,z)$:
\begin{equation}
\label{e4}\mbox{(a)}\quad\frac{{\rm d}\Um}{{\rm d} t}   = 
\frac{1}{2 \R} \left[ \frac{\partial^2 \phi_x }{\partial y^2}\right]_{y=-1}^{y=+1}+\frac{2}{\R} 
\,,\hspace{3em} \mbox{(b)}\quad
\frac{{\rm d}\Wm}{{\rm d}t}   = 
\frac{1}{2 \R} \left[ \frac{\partial^2 \phi_z}{\partial y^2}  \right]_{y=-1}^{y=+1}\,,
\end{equation}
where the term $2/\R$ in (\ref{e4}a) accounts for the constant 
streamwise driving.

In addition to periodic boundary conditions applied 
at distances $L_x$ and $L_z$ 
to $\omega_y$ and $u_y$, the no-slip boundary conditions at the walls 
and the boundary conditions relative to $\phi_x$ and $\phi_z$ are: 
\begin{equation}
\label{e-bc}
\left. u_y\right|_{y=\pm 1}
=\left. \frac{\partial u_y}{\partial y}\right|_{y=\pm 1}=\left. \omega_y\right|_{y=\pm 1}=
\left. \frac{\partial \phi_x}{\partial y} \right|_{y=\pm 1} = \left. \frac{\partial \phi_z}{\partial y} \right|_{y=\pm 1}=0,~
\left. \phi_x \right|_{y=\pm 1}=\pm \Um,~
\left. \phi_z \right|_{y=\pm 1}=\pm \Wm.
\end{equation}
The set of equations \eqref{e2}-\eqref{e4} with boundary conditions (\ref{e-bc}) are numerically integrated as follows:\\
-- Spatial treatment of \eqref{e2} uses Fourier series in streamwise 
and spanwise directions, $x$ and $z$, respectively.
In view of numerical accuracy and efficiency, the wall-normal 
dependence of \eqref{e2} is dealt with combinations of Chebyshev 
polynomials satisfying the  
homogeneous boundary conditions at $y=\pm 1$ \cite{S95}.
The equations for $\omega_y$ and $u_y$ are solved using
the following expansions:
\begin{gather}
\label{me1}
\omega_y = \sum_{l=0}^{L-2} \sum_{m=-M}^M  \sum_{n=-N}^N \hat{\omega}^{lmn} 
\left(T_{l+2} (y)- T_l (y)\right) \exp\left[ 
{\rm i}\mbox{$\frac{2 \pi}{L_x}$}mx+{\rm i}\mbox{$\frac{2 \pi}{L_z}$} nz \right], \\
\label{me2}
u_y = \sum_{l=0}^{L-4} \sum_{m=-M}^M  \sum_{n=-N}^N \hat{u}^{lmn}  
\left( T_{l+4} (y)- \frac{2l+4}{l+3} T_{l+2} (y) + \frac{l+1}{l+3} T_l (y) \right)
\exp \left[{\rm i}\mbox{$\frac{2 \pi}{L_x}$}mx+{\rm i}\mbox{$\frac{2 \pi}{L_z}$} nz \right],
\end{gather}
where $T_l$ is the Chebyshev polynomial of degree $l$.\\
-- Similarly, the auxiliary fields $\phi_x$ and $\phi_z$ in \eqref{e3} 
are expanded as
\begin{gather}
\label{me3}
\phi_x =  \sum_{l=0}^{L-4}   \hat{\phi}^{l}_x
\left( T_{l+4} (y)- \frac{2l+4}{l+3} T_{l+2} (y) + \frac{l+1}{l+3} T_l (y) \right) 
+\frac{\Um}{2}y(3-y^2)\,, \\
\label{me4}
\phi_z =  \sum_{l=0}^{L-4}   \hat{\phi}^{l}_z
\left( T_{l+4} (y)- \frac{2l+4}{l+3} T_{l+2} (y) + \frac{l+1}{l+3} T_l (y) \right)
+\frac{\Wm}{2}y(3-y^2)\,,
\end{gather}
where the last terms in (\ref{me3},\ref{me4}) follow from the 
boundary-condition homogenization technique \cite{B01}; these terms are 
the polynomials of lowest degree satisfying all boundary conditions 
for $\phi_x$ and $\phi_z$, respectively.\\
-- A conventional Galerkin method is developed by taking the inner product 
of the basis functions and the evolution equations, 
yielding ordinary differential equations for each coefficient in the expansions.
Periodic boundary conditions are imposed in the streamwise and spanwise 
directions at distances $L_x=500$ and $L_z=250$.
Maximum Fourier wavenumbers are $M=N=767$ and the maximum degree 
for Chebyshev polynomials in the $y$ direction is $L=31$.
Aliasing errors involved in the evaluation of the quadratic nonlinear terms 
are removed in all directions using the 2/3 rule \cite{B01} in the $x$ and $z$
directions, and computing all coefficients of 
Chebyshev polynomials up to degree $2L$ in direction $y$.
The evaluation of nonlinear terms then involves $(N_x,N_y,N_z)=(2304,64,2304)$ modes.
\revv{For larger and smaller domain sizes $(L_x,L_z)=(1000,500)$ and $(250,125)$ used to check size effects,  mode numbers are taken in the same proportions.}
This spatial resolution has been found appropriate from the comparison 
with other numerical simulations and a parallel study of plane Couette flow 
driven by counter-translating plates rather than by a constant body force.
The equations are numerically time-integrated in a standard way using 
a second-order method, Crank--Nicolson for the viscous terms 
and Heun method for the other terms, with time-increment 
$\delta t=0.04$.
\medskip

Other works may solve the flow using different definitions or scalings:\\
-- In the literature, reference is often made to the mean flow rate.
Here the mean streamwise bulk velocity is a measured quantity 
that can characterize the flow regime upon time averaging, 
see Fig.~\rev{3(a)} displaying $\average{\Um}$ as a function of $\R$.
From it we can define a Reynolds number $\Rm$ of practical use, 
though not a control parameter, as $\Rm= 1.5\, \R \, \average{\Um}$.
Once our scaling is adopted for the velocity, $\Rm$ is then 
understood as the Reynolds number built using the center-plane 
velocity of a parabolic flow profile with the same mean velocity.\\
-- Another popular choice is by using so-called {\it wall units}:
The friction velocity is defined by $U_\tau^2 = \tau_{\rm w} = 
\nu \average{|\partial_y u_x|}_{\rm wall}$.
It is obtained by averaging (\ref{e1}b) as $U_\tau^2=2/\R$ 
in our unit system; then,
\begin{equation}
\R_\tau=\R\, U_\tau=\sqrt{2\R}.
\label{Retau}
\end{equation}
The friction coefficient is traditionally defined as $\Cf= \tau_{\rm w}/\frac12 \average{\Um}^2$.
In our approach, its expression directly comes from $\tau_{\rm w}=U_\tau^2
=2/\R$ to give $\Cf=4/(\R \average{\Um}^2)$.
Taking the inner product between $\bm u$ and (2b) and 
averaging the equation then relates $\Cf$ to the mean dissipation rate $\average{\epsilon_{\rm m}}$.
Expressed using $\average{\Um}$ as the velocity unit, the volume-averaged dissipation rate
$\epsilon_{\rm m}(t)=1.5 (\Rm\average{\Um}^2)^{-1}\left[{\mathcal V}^{-1}
\int_{\mathcal V}  (\nabla {\bm u} \cdot \nabla {\bm u})\,{\rm d} {\mathcal V}\right]$ 
is computed to give:
\begin{equation}
\label{epsilonm}
\average{\epsilon_{\rm m}}=\Cf/2.
\end{equation}

\section{Measurement of turbulent fraction\label{AppB}}

Local laminar and turbulent states are discriminated according to a traditional method of computer image treatment called ``moment-preserving thresholding''~\cite{T85}.
The principle is to separate pixels in an image into several classes, using thresholds defined in a systematic way from the statistics of the pixels'  values, typically gray levels, rather than from some empirically-defined rule.
On general grounds, the thresholds are chosen so that the first few moments of the histograms of pixel values are preserved.
In the present case, two classes distinguished through a single threshold, are needed: ``turbulent$\,\equiv\,$above'' and ``laminar$\,\equiv\,$below''.
Normalized histograms are obtained from the distribution of the observable, here simply called $u$, at all points in the domain, 
say $p(u)$. Its moments $m_k=\sum_u u^k p(u)$, $k=1,2,\dots$ are next computed.
Unknown reduced variables $\tilde u_\ell$ with probability $p_\ell$ for laminar local states and $\tilde u_t$ with probability $p_t=1-p_\ell$ for turbulent ones are then determined using three equations for the three lowest order moments \cite{T85}:
${\tilde u_\ell}^k p_\ell + {\tilde u_t}^k p_t=m_k$, $k=\{1,2,3\}$.
(Multi-level thresholding would require more moments and,  if the set of pixels were to be divided into $M$ classes; it is easily be seen that $2M-1$ moments would then be required.)
The reduced representation $(\tilde u_\ell,\tilde u_t)$ is thus the best two-level representation of the original field, while probability $p_t$ corresponds to our evaluation of  the turbulent fraction $F_{\rm t}$. 

\begin{figure}
\begin{minipage}{0.3\hsize}
\begin{center}
\vspace{1mm}
raw\\[0.5ex]
\includegraphics[width=0.96\textwidth]{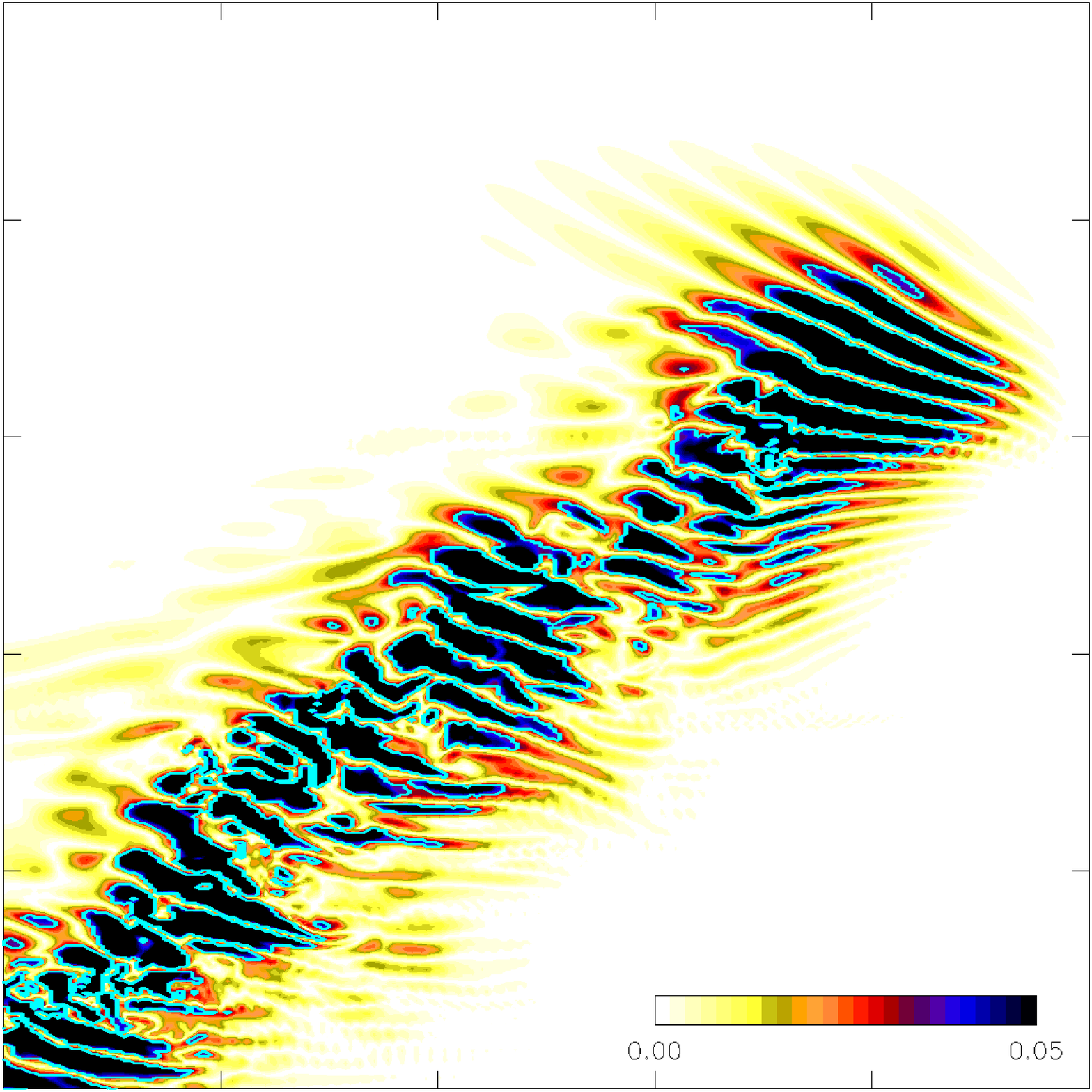}\\
\end{center}
\end{minipage}
\begin{minipage}{0.3\hsize}
\begin{center}
$w=12\delta$\\[0.5ex]
\includegraphics[width=0.96\textwidth]{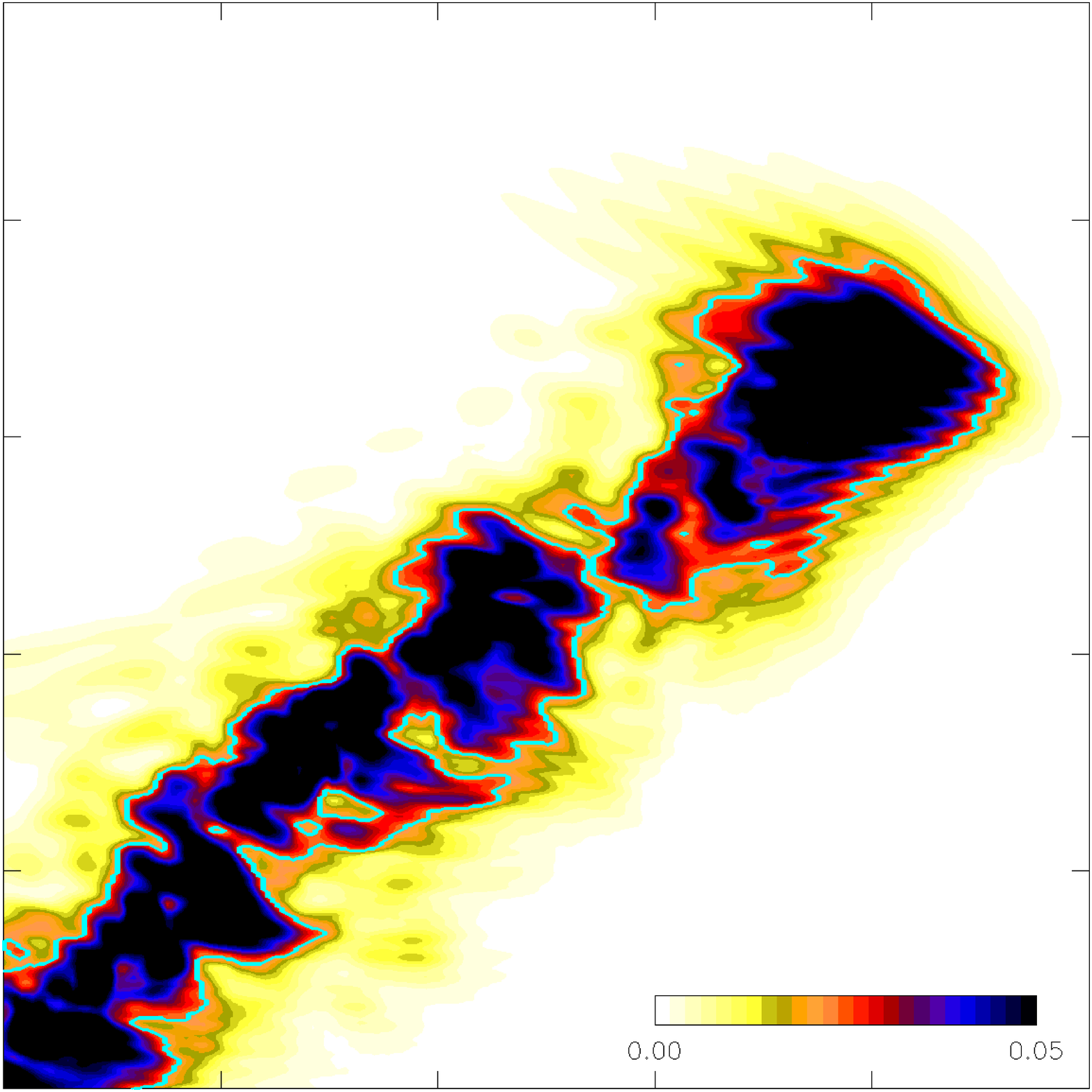}
\end{center}
\end{minipage}
\begin{minipage}{0.3\hsize}
\begin{center}
$w=40\delta$\\[0.5ex]
\includegraphics[width=0.96\textwidth]{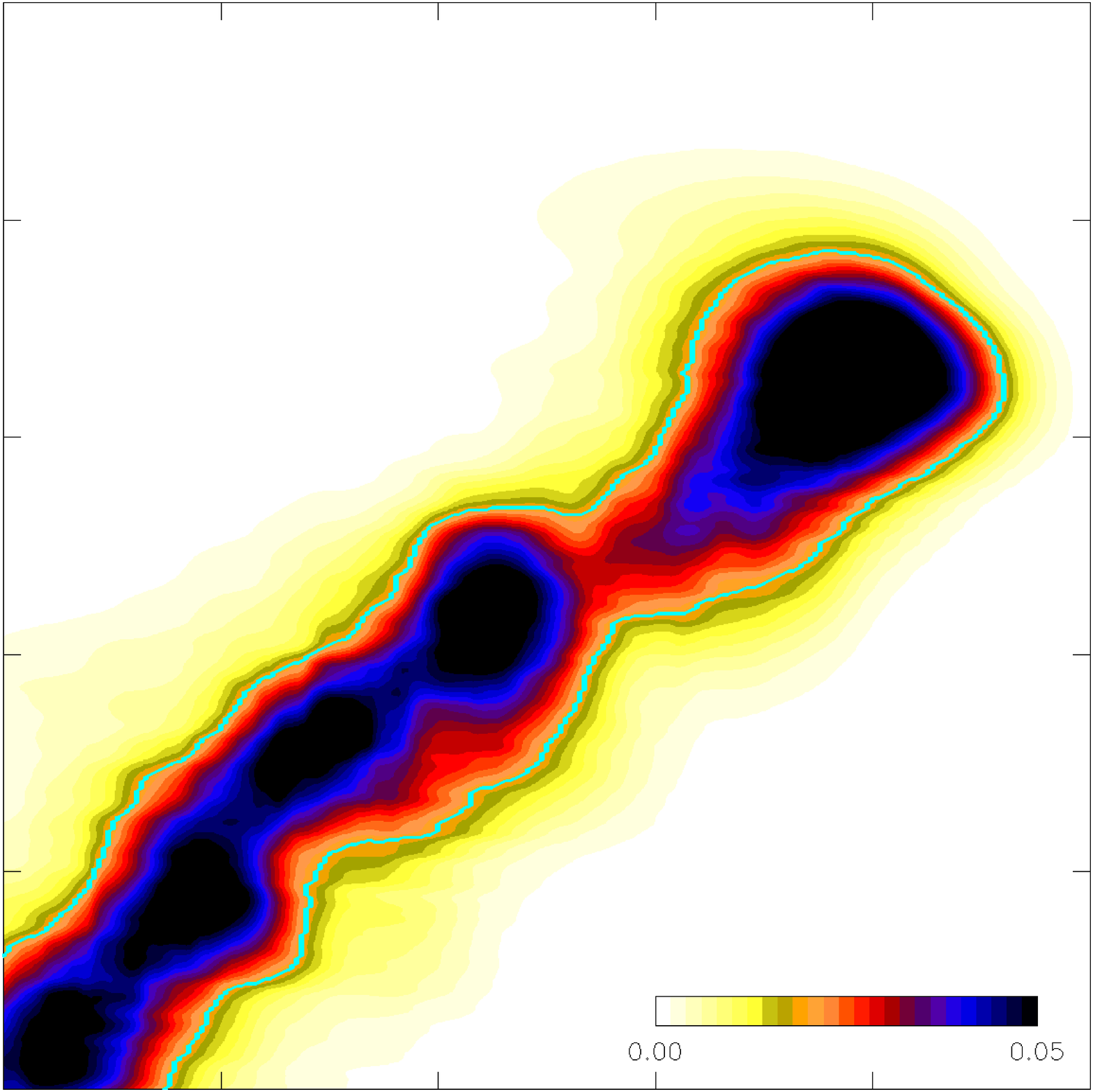}
\end{center}
\end{minipage}
\caption{Color-level illustration of the DAH of the LTB shown in Fig.~\ref{Sltb} using $|u_y(x,0,z)|$. 
The size of the domain displayed is $50\times 50 \simeq 460\delta\times460\delta$. \rev{Flow direction is from left to right.}
Raw data (left) and after box-filtering with the width 
$w=12\delta$ (routinely used here) and $w=40\delta$.
Light-blue lines mark the boundaries between turbulent and laminar regions as determined by the moment-preserving thresholding method.}
\label{Swth}
\end{figure}

\begin{figure}
\begin{minipage}{0.49\textwidth}
\begin{center}
$\R=1200$ \\[0.5ex] 
\includegraphics[width=0.95\textwidth]{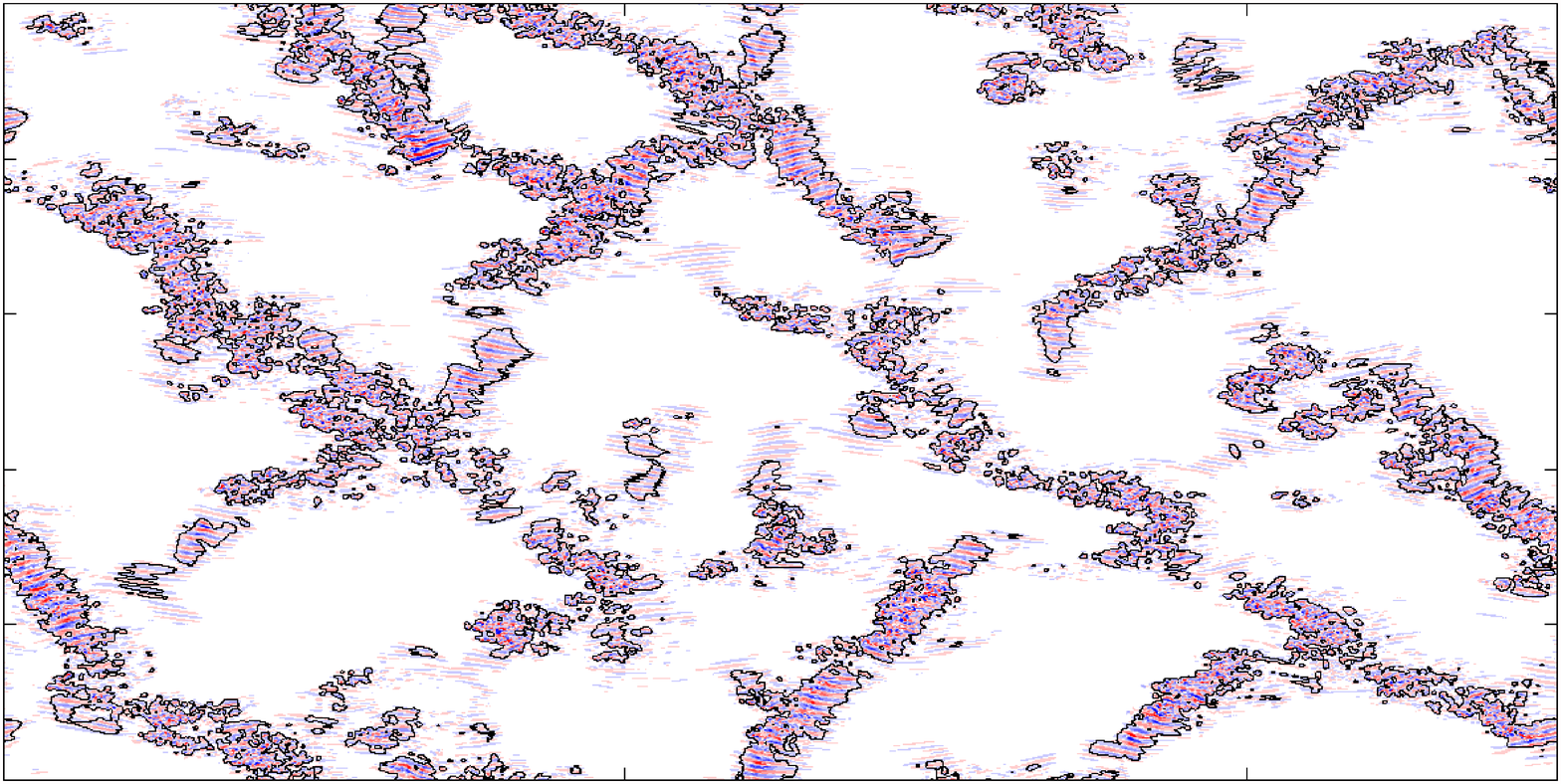}
\end{center}
\end{minipage}
\hfill
\begin{minipage}{0.49\textwidth}
\begin{center}
$\R=1800$ \\[0.5ex] 
\includegraphics[width=0.95\textwidth]{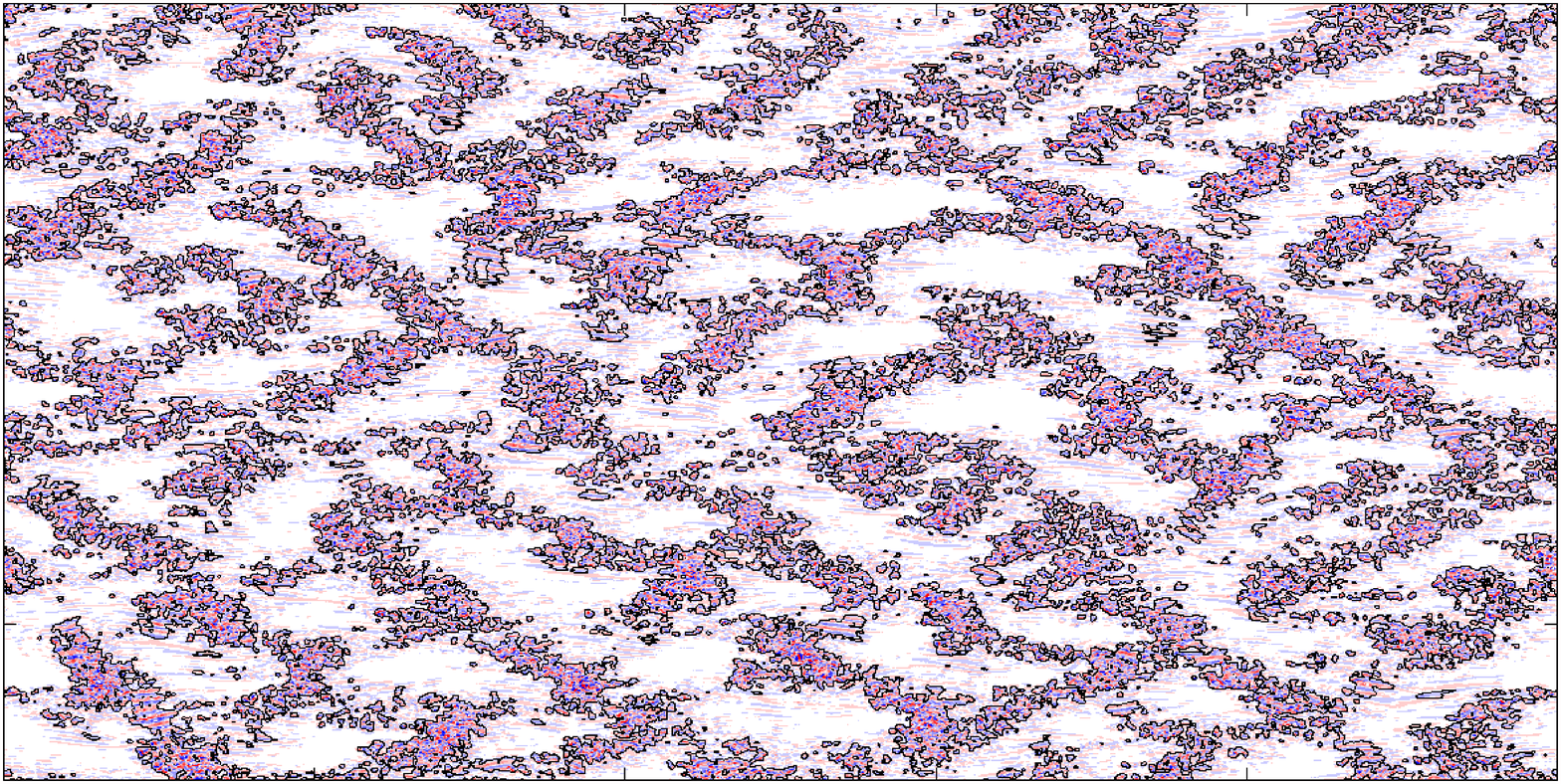}
\end{center}
\end{minipage}
\caption{\label{Sthrre}  Moment-preserving thresholding of snapshots at $1200$ and $1800$ in Fig.~1(c,d) under box-filtering with $w=12\delta$. \rev{Flow direction is from left to right. Domain size is $500\times 250$.}}
\end{figure}

The observable  $u$ of interest is here the box-filtered value of  $|u_y(x,0,z)|$ and a single parameter remains, the width $w$ of  the squares $w\times w$ over which $|u_y(x,0,z)|$ is averaged.
In order to visualize the result of the procedure, the laminar/turbulent cut-off $u_*$ has to be determined.
This is done by expressing the condition defining $p_t$ explicitly, that is:
$p_t=\#(u_i > u_*)/(N_xN_z)$, where $\#(u_i > u_*)$
is the number of  lattice nodes considered as turbulent. 
The most faithful reproduction of the contour of turbulent domains requires that $w$ should be large enough to damp out irrelevant small-scale modulations seen in Fig.~\ref{Swth}, ``raw''.
The other panels in Fig. \ref{Swth} illustrate the output of the procedure for different values of $w$. 
In the figure, $\delta$ stands for the spanwise grid spacing $\delta = L_z/N_z \simeq 0.109$.
As the filter size becomes larger, spurious laminar regions disappear.
A thin line corresponding to the cut-off condition $u=u_*$ delineates the turbulent region obtained with our thresholding method.
Figure~\ref{Sthrre} displays the two snapshots in Fig.~\ref{snap} belonging to the $\R$ range studied in \S\ref{S4} using $w=12\delta$.

\newpage

\section{Supplemental video}

\subsection*{Re=725}

One-sided LTB regime at $\R=725$ [Fig. 4(a) and (b)].
Splittings and collisions between LTBs propagating in the same direction
are observed but without any trace of a transversal splitting. 
Wall-normal velocity field on the mid-plane $u_y(x,0,z,t)$ is displayed 
in a co-moving reference frame with velocity $0.8$ in stream-wise direction 
(to the right) for better visibility.

\subsection*{Re=850}

One-sided LTB regime with rare transversal splittings at $\R=850$ 
[Fig. 1(a)].
The displayed quantity and velocity of the reference frame are the 
same as in Supplemental Video Re=725.

\subsection*{Re=900}

One-sided LTB regime with frequent transversal splittings at $\R=900$ 
[Fig. 4(c) and (d)].
The displayed quantity and velocity of the reference frame are the same 
as in Supplemental Video Re=725.

\subsection*{Re=1050}

Two-sided LTB regime at $\R=1050$, slightly above 
the onset at $\R_2\simeq1011$[Fig. 1(b)].
The displayed quantity is the same as in Supplemental Video Re=725.
The velocity of the reference frame is $0.7$. 

\subsection*{Re=1200}

Strongly intermittent loose continuous network of LTBs at $\R=1200$ 
[Fig. 1(c)].
The displayed quantity is the same as in Supplemental Video Re=725.
The velocity of the reference frame is $0.6$. 

\subsection*{Re=1800}

Weakly intermittent loose banded pattern at $\R=1800$ 
[Fig. 1(d)].
The displayed quantity is the same as in Supplemental Video Re=725.
The velocity of the reference frame is $0.45$. 

\subsection*{Re=3000}

Tight banded pattern at $\R=3000$ [Fig. 1(e)].
The displayed quantity is the same as in Supplemental Video Re=725.
The velocity of the reference frame is $0.35$.

\subsection*{One-sided regime in a larger domain at Re=900} 

Everything is same as in Supplemental Video Re=900 except for the horizontal domain size, which is 1000 and 500 in 
streamwise and spanwise directions respectively. 
The initial transient state is included to show the transition from 
randomly distributed LTBs of both directions to the final  one-sided state.

\subsection*{Two-sided regime in a larger domain at Re=1050}
Everything is same as in Supplemental Video Re=1050 except for  
the horizontal domain size, which is 1000 and 500 in 
streamwise and spanwise directions respectively.


\begin{thebibliography}{58}
\bibitem [{NRm()}]{NRm}%
  \BibitemOpen
  \href@noop {} {}\bibinfo {note} {Finite-size effects linked to this
  methodological difference presumably roots a systematic discrepancy between
  our definition of the mean-flow based {R}eynolds number and the one
  in~\cite{ST16}, leaving us unable to match results about DP thresholds
  quantitatively.}\BibitemShut {Stop}%
\end{thebibliography}
%

\end{document}